\documentclass[structabstract]{aa}  
\usepackage{float}                                   
\usepackage{amsmath}
\usepackage{natbib}
\usepackage{lscape}
\usepackage{fixltx2e}
\usepackage{rotating}
\usepackage{graphicx}
\usepackage[varg]{txfonts}
\usepackage{xcolor}
\usepackage{hyperref}
\usepackage{mathrsfs}
\usepackage{bbding}
\usepackage{verbatim}
\usepackage{amssymb}

\def\sun{\odot}
 
 \def\mso{\,\mathrm{M}_\odot}
 \def\rso{\,{\rm R}_\odot}
 \def\lso{\,{\rm L}_\odot}

 \def\llso{\log\, L/{\rm L}_\odot \,}
 \def\simle{\mathrel{\hbox{\rlap{\hbox{\lower4pt\hbox{$\sim$}}}\hbox{$<$}}}}
 \def\simgr{\mathrel{\hbox{\rlap{\hbox{\lower4pt\hbox{$\sim$}}}\hbox{$>$}}}}

\newcommand{\deb}[1]{\mathrm{#1}}

\begin{document}

   \title{Massive main sequence stars evolving at the Eddington limit}


   \author{D. Sanyal
          \thanks{e-mail: dsanyal@astro.uni-bonn.de} 
          \and
          L. Grassitelli 
           \and
          N. Langer   
          \and
          J. M. Bestenlehner\thanks{Present address: Max-Planck-Institut f\"ur Astronomie, K\"onigstuhl 17, D-69117 Heidelberg, Germany} 
	      }

   \institute{Argelander-Insitut f\"ur Astronomie, Universit\"at Bonn, Auf
              dem H\"ugel 71, 53121 Bonn, Germany
             }

   \date{Received February 2015}

 
  \abstract
   {Massive stars play a vital role in the Universe. However, their evolution even 
   on the main sequence is not yet well understood.
   }
   {Due to the steep mass-luminosity relation, massive main sequence stars become
   extremely luminous. This brings their envelopes very close to the Eddington limit. 
   We are analysing stellar evolutionary models in which the Eddington limit is reached
   and exceeded, and explore the rich diversity of physical phenomena 
   which take place in their envelopes, and we investigate their observational consequences.}
   {We use the published grids of detailed stellar models by Brott et al. (2011) 
   and K\"{o}hler et al. (2015), computed with a state-of-the-art one-dimensional 
   hydrodynamic stellar evolution code using LMC composition, to investigate 
   the envelope properties of core hydrogen burning massive stars.}
   {We find that at the stellar surface, the Eddington limit is almost never reached, 
   even for stars up to $500\mso$. When we define an appropriate Eddington limit 
   locally in the stellar envelope, we can show that most stars more massive 
   than $\sim 40\mso$ actually exceed this limit, in particular in the partial 
   ionization zones of iron, helium or hydrogen. While most models adjust their 
   structure such that the local Eddington limit is exceeded at most by a few per cent, 
   our most extreme models do so by a factor of more than seven. We find that the 
   local violation of the Eddington limit has severe consequences for the envelope structure, 
   as it leads to envelope inflation, convection, density inversions and possibly to pulsations. 
   We find that all models with luminosities higher than $4\times 10^{5}\lso$, i.e. stars above 
   $\sim 40\mso$ show inflation, with a radius increase of up to a factor of about 40. 
   We find that the hot edge of the S\,Dor variability region coincides with a line beyond which 
   our models are inflated by more than a factor of two, indicating a possible connection 
   between S\,Dor variability and inflation. \normalfont{Furthermore, our coolest models show 
   highly inflated envelopes with masses of up to several solar masses, 
   and appear to be candidates to produce major LBV eruptions.}
} 
   {Our models show that the Eddington limit is expected to be reached in all stars above $\sim 40\mso$ in the LMC,
    and by even lower mass stars in the Galaxy, or in close binaries or rapid rotators. 
    While our results do not support the idea of a direct 
    super-Eddington wind driven by continuum photons, 
    the consequences of the Eddington limit in the form of 
    inflation, pulsations and possibly eruptions may well give rise to
    a significant enhancement of the the time averaged mass loss rate.
     }
     

    \authorrunning{D. Sanyal et al.}
   \maketitle
%

\section{Introduction}\label{sec:intro}

Massive stars are powerful engines and strongly affect the evolution of star forming galaxies
throughout cosmic time \citep{Bresolin_2008}. In particular the most massive ones produce
copious amounts of ionising photons \citep{doran_2013}, 
emit powerful stellar winds \citep[e.g.,][]{kudritzki2000,smith_2014} and in their final explosions 
are suspected to produce the most energetic and spectacular stellar explosions, as
pair-instability supernovae \citep{kozyreva_2014}, 
superluminous supernovae \citep{gal-yam2009}, 
and long-duration gamma-ray bursts \citep{larsson2007,raskin2008}.

Massive main sequence stars, which we understand here as those which undergo core hydrogen burning,
have a much higher luminosity than the Sun, as they
are known to obey a simple mass-luminosity relation, $L\sim M^{\alpha}$, with $\alpha > 1$.
However, whereas this relation is very steep near the Solar mass ($\alpha\simeq 5$),
it is shown in \citet{kw90} that $\alpha \rightarrow 1$ for $M \rightarrow \infty$.
Indeed, \citet{koehler2015} find $\alpha\simeq 1.1$ for $M=500\mso$.

Since the Eddington factor is proportional to $L/M$, it is debated in the literature 
whether main sequence stars of higher and higher initial mass eventually
reach the Eddington-limit \citep{langer97,crowther2010, maeder_2012}. 
The answer is clearly: yes, they do.
Even when only electron scattering is considered as a source of 
radiative opacity, the Eddington-limit corresponds to a luminosity-to-mass ratio
of $\mathcal{R} :=\log \left( {L\over \lso}/{M\over \mso}\right) \simeq 4.6$ \citep{langer2014} for hot stars
with a solar helium abundance. This is extremely close to the $\mathcal{R}$-values 
obtained for models of supermassive stars, where this ratio is nearly mass-independent
\citep{fuller_1986,kato86}. In fact,
\citet{kato86} showed that zero-age main sequence models computed only with 
electron scattering opacity do reach the Eddington limit at a mass of about $\sim 10^5\mso$.

Whether supermassive stars exist is an open question. Also the mass limit of
ordinary stars is presently uncertain \citep{schneider_2014}. 
However, there is ample evidence for stars with initial masses well
above $100\mso$ in the local Universe. 
A number of close binary stars have been found with component initial masses above $100\mso$
\citep{schnurr_2008,schnurr_2009,taylor_2011,sana_2013}. \citet{crowther2010}
proposed initial masses of up to $300\mso$ for several stars in the Large Magellanic Cloud (LMC),
based on their luminosities. \citet{bestenlehner2014} identified 
more than a dozen stars more massive than $100\mso$ from the sample of $\sim 1000$ OB\,stars
near 30\,Doradus, which are analysed in the frame of VLT-Flames Tarantula Survey \citep{evans2011}.
The hydrogen-rich stars among them have measured  $\mathcal{R}$-values of up to 4.3.
In hot stars with finite metallicity, the ion opacities can easily exceed the electron
scattering opacity \citep{ir96}. It is thus to be expected that the true Eddington limit,
which accounts for all opacity sources, is located at $\mathcal{R}$-values of 4.3 or below. 
This implies that these stars should in fact also have reached, or exceeded their Eddington limit.   

In this paper, we explore the question of massive main sequence stars reaching, or exceeding 
the Eddington limit from the theoretical side.
We show by means of detailed stellar models as described in Section\,\ref{sec:modelling} 
that even all stars more massive than $\sim 40\mso$ are found to reach the Eddington limit. 
In Section\,\ref{sec:edd_limit}, we demonstrate the need to properly 
define a local Eddington factor in the stellar
interior, which we then use in Section\,\ref{sec:inflation} to 
show that when it exceeds the critical value of one, 
the stellar envelope becomes inflated. We show further in Section\,\ref{sec:dens_inv} 
and\,\ref{sec:convective_velocity} that super-Eddington conditions
can lead to density inversions, and induce convection. We compare our results to
previous studies in Section\,\ref{sec:comparison_lit}, and 
relate them to observations in Section\,\ref{sec:comparison_obs},
before summarising our conclusions in Section\,\ref{sec:conclusions}.

\begin{figure}
\centering
 \resizebox{\hsize}{!}{\includegraphics[angle=-90]{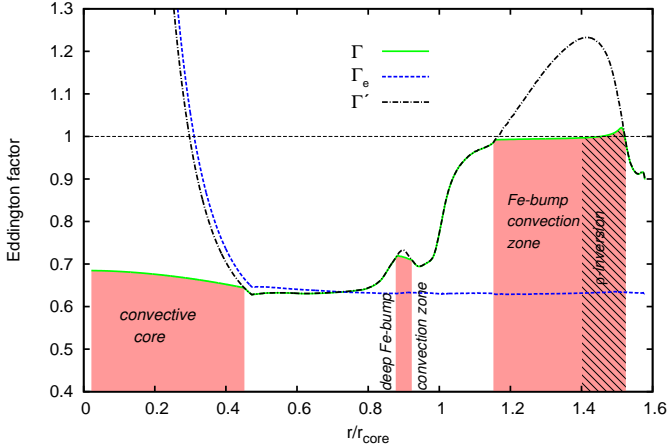}}
   \caption[]{
   The different Eddington factors inside a $285\,\deb{M_{\sun}}$, 
   non-rotating, main sequence model with 
   $\log L/L_{\sun}=6.8$ and $\deb{T_{eff}=46600\,K}$ (cf. Fig.\,\ref{fig:inflation_eg} and Appendix \ref{app:inflations_eg}). 
   The shaded areas mark the different convection zones and the hatched 
   area marks the region with a density inversion.  
   The radius of the un-inflated core is denoted as 
   $r_{\deb{core}}$ (defined in Sect.\, \ref{sec:inflation}). The black dashed horizontal line  
   is drawn at $\Gamma=1$ for convenience. 
   }
\label{fig:gamma_eg}
\end{figure}

\begin{figure*}
\begin{center}
\resizebox{\textwidth}{!}{\includegraphics[angle=-90]{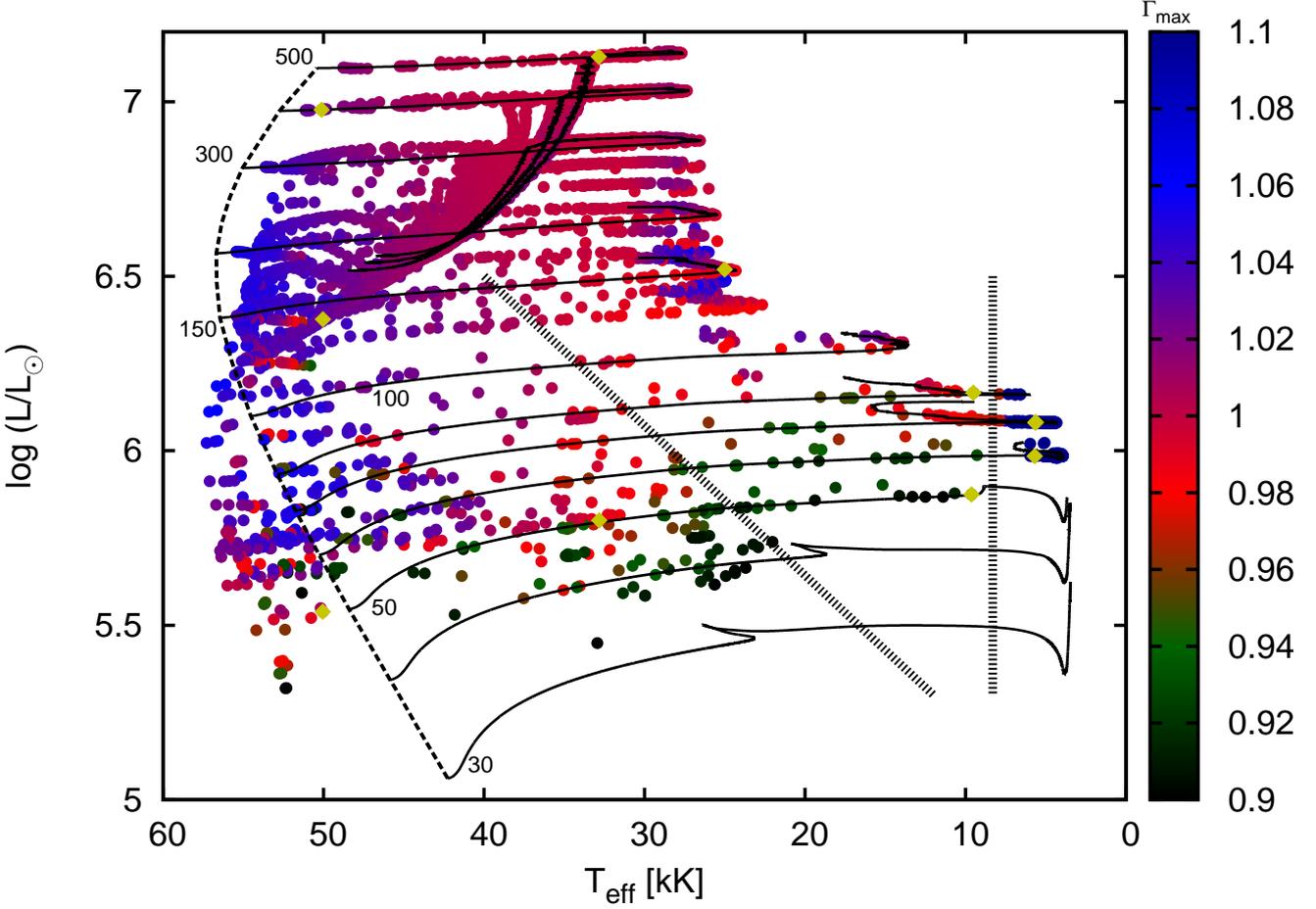}}
   \caption[]{
     Positions of the analysed stellar models with $\Gamma_{\mathrm{max}}>0.9$ 
     in the Hertzsprung-Russell diagram (colored dots).
     Models with  $\Gamma_{\mathrm{max}}>1.1$ are colored dark-blue. 
     The solid lines show the evolutionary tracks of non-rotating 
     stellar models \citep{koehler2015}. The initial masses 
     are marked in units of solar mass. The dashed line corresponds to the 
     zero-age main sequence of the non-rotating models. The 
     hot and the cool edges of the S\,Dor instability strip from \citep{smith2004} 
     are indicated with thick dotted lines. 
     The  interior structures of the models marked with yellow diamonds  
     are shown in Appendix \ref{app:example_models}.} 
  \label{fig:gamma_HR}
\end{center}
\end{figure*}
\section{Stellar models}\label{sec:modelling}

The grids of stellar models used for the 
present study have been published in \citet{ines2011} and \citet{koehler2015}. 
In this paper, we consider only the core hydrogen burning models computed with LMC metallicity.
Each stellar evolution sequence computed by \citet{ines2011} and \citet{koehler2015} consists
of typically 2000 individual stellar models. However, the full amount of data defining a stellar model
is only stored for a few dozen time points per sequence, in non-regular intervals.
It is those stored models which are analysed here. This scheme has the disadvantage that 
the density of models in the investigated parameter space is not always as high
as it should be ideally. Still, as shown below, it allows for a thorough sampling of
the considered parameter space, and it is fully consistent
with the results already published. 

The stellar models were computed with a state-of-the-art
one-dimensional hydrodynamic implicit Lagrangian
code (BEC) which incorporates latest input physics
\citep[for details, see][and references therein]{bra97,yln06,ines2011,koehler2015}.
Convection was treated as in the standard 
non-adiabatic mixing length approach \citep{MLT1958,kw90} and a 
mixing length parameter of $\alpha=l/H_{\rm p}=1.5$ \citep{lan91} 
was adopted, with $l$ and $H_{p}$ being the mixing length 
and the pressure scale height respectively. 
This value of the mixing length parameter does lead to a good representation 
of the Sun \citep{suijs2008}, whereas its calibration to 
multi-dimensional hydrodynamic models shows
that it tends to decrease towards lower gravities \citep{trampedach2014,magic2015}.
The convective velocities were limited to the local value of the adiabatic
sound speed. The contribution of turbulent pressure 
\citep{deJager1984} was neglected,  
since it is not expected to be important 
in determining the stellar hydrostatic structure \citep[]{stothers_2003}. 
Indeed, our recent study which includes turbulent pressure 
(Grassitelli et al. in preparation) shows that e.g. for an 
$80\mso$ evolutionary sequence, the stellar radius is increased over that of
models without turbulent pressure by at most a few per cents at any time during  
its main-sequence evolution. 
Rotational mixing of chemical elements following \citet{heger2000} and transport of angular momentum 
by magnetic fields due to the Spruit-Taylor dynamo were also included \citep{spruit02}. 
The efficiency parameters $f_c$ and $f_{\mu}$ for rotational mixing were set to $0.0228$ and 
$0.1$ respectively \citep{ines2011}. Radiative opacities were 
interpolated from the OPAL tables \citep{ir96}. The opacity enhancement 
due to Fe-group elements at $T \sim 200$ kK plays a vital role in 
determining the envelope structure in our stellar models. We note that
even though flux-mean opacities are appropriate to study the momentum balance 
near the stellar photosphere, we only consider the Rosseland mean opacities in the
following, which are thought to behave very similarly to the flux-mean opacities 
especially at an optical depth larger than one.

The outer boundary condition of the stellar models corresponds to a 
plane-parallel gray atmosphere model 
on top of the photosphere. In other words, 
the effective temperature was used as a 
boundary condition at a Rosseland optical depth of $2/3$. 
The adopted stellar wind mass loss recipe 
does lead to small but finite
outflow velocities in the outermost 
layers, which induces a 
slight deviation from hydrostatic equilibrium.  

The mass loss prescription from \citet{vink2000,vink2001} 
was employed to account for the winds of 
O- and B-type stars. Moreover, parameterized mass loss 
rates from \citet{nieuwenhuijzen1990} were used on the cooler side 
of the bi-stability jump, i.e. at effective temperatures less than 22000 K, if the 
\citet{nieuwenhuijzen1990} mass loss rate exceeded that of \citet{vink2000,vink2001}. 
Wolf-Rayet (WR) type mass loss was accounted for using the empirical 
prescription from \citet{hamann95} divided by a factor of 10 \citep{yln06}, when the surface helium 
mass fraction became greater than $70\%$. 

Evolutionary sequences of massive stars,  
with and without rotation, were computed up to 
an initial mass of $500\,\mathrm{M_{\sun}}$, starting with LMC composition. 
The initial mass fractions of hydrogen, helium and metals 
were taken to be 0.7391, 0.2562 and 0.0047 respectively, 
in accordance with the observations of young massive stars in the LMC \citep{ines2011}.

\section{The Eddington Limit}\label{sec:edd_limit}

The Eddington limit refers to the condition 
where the outward radiative acceleration in a star 
balances the inward gravity, in hydrostatic equilibrium. 
It is a concept which is thought to apply at the 
stellar surface, in the sense that
if the Eddington limit is exceeded, a mass outflow 
should arise \citep{eddington_1926,owocki2004}.
If we denote the gravity as $g=GM/r^2$ and the 
radiative acceleration mediated through the electron scattering opacity
as $g_{\deb{rad}}=\kappa_{\deb e} L/4\pi r^2$, then 
the classical Eddington factor $\Gamma_{\deb e}$ is defined as 
\begin{equation}\label{eq:gamma_e}
 \Gamma_{\deb e}:=\frac{L}{L_{\mathrm{Edd}}}=\frac{g_{\deb{rad}}}{g}=\frac{\kappa_e L}{4\pi c G M},
\end{equation}
where $L, M$ and $\kappa_{\deb e}$ are the luminosity, 
mass and electron-scattering opacity respectively, 
with the physical constants having their usual meaning. The classical 
Eddington parameter $\Gamma_{\deb e}$ therefore does not 
depend on the radius $r$ as the inverse $r^2$ scaling in 
both $g_{\deb{rad}}$ and $g$ cancel out.
Whereas $\Gamma_{\deb e}$ is often convenient to consider, it 
provides a sufficient instability criterion to stars but not a necessary one, because 
usually the true opacity does exceed the the electron scattering opacity significantly 
and also contributes to the radiative force.

As it turns out below, even when the Rosseland mean opacities are used, 
the models analysed in this paper practically never
reach the Eddington limit at their surface. Therefore,
we instead consider the Eddington factor in the stellar interior as
\begin{equation}\label{eq:gamma}
 \Gamma'(r):=\frac{L(r)}{L_{\mathrm{Edd}}(r)}=\frac{\kappa(r)L(r)}{4\pi c G M(r)},
\end{equation}
where $M(r)$ is the Lagrangian mass coordinate, 
$\kappa(r)$ is 
the Rosseland mean opacity and $L(r)$ is the 
local luminosity \citep{langer97}. 
However, $\Gamma'(r)> 1$ also does not provide a stability limit 
in the stellar interior because the stellar layers turn convectively 
unstable following Schwarzschild's criterion 
when $\Gamma'(r)\rightarrow 1$ \citep{joss73,langer97}. 
As the luminosity transported by convection does not
contribute to the radiative force, we subtract the convective luminosity 
in the above expression and redefine the Eddington factor as 
\begin{equation}\label{eq:gamma_new}
 \Gamma(r):=\frac{L_{\mathrm{rad}}(r)}{L_{\mathrm{Edd}}(r)}=\frac{\kappa(r)(L(r)-L_{\mathrm{conv}}(r))}{4\pi c G M(r)}.
\end{equation}
For example, near the stellar core where convective energy transport 
is highly efficient, $ \Gamma(r)$ stays well below unity in spite of 
$\Gamma'(r)\gg 1$ and no instability, i.e. departure from 
hydrostatic equilibrium, occurs (see Fig.\, \ref{fig:gamma_eg}). In the rest 
of the paper we will refer to $\Gamma(r)$ as the Eddington factor 
unless explicitly specified otherwise.

Even with this definition, a super-Eddington layer 
inside a star does not necessarily lead 
to a departure from hydrostatic equilibrium or a sustained mass outflow. 
In the outer envelopes of massive stars non-adiabatic 
conditions prevail and convective energy 
transport is highly inefficient which pushes $\Gamma(r)$ 
close to (or above) one. 
We find that the stellar models counteract such a super-Eddington luminosity 
by developing a positive gas pressure gradient, 
thus restoring hydrostatic equilibrium \citep{langer97,asplund_1998}. 
In such situations the canonical definition 
of $L_{\mathrm{edd}}$ being the maximum sustainable radiative
luminosity locally in the stellar interior (in hydrostatic equilibrium)  
breaks down and loses its significance. As we shall see below 
the radiative luminosity beneath the photosphere can be up to a 
few times the Eddington luminosity.

In Fig.\, \ref{fig:gamma_eg}, the behavior of  
$\Gamma $ and $\Gamma'$ is shown
along with the electron-scattering Eddington factor $\Gamma_\deb{e}$ 
in a $285\, \deb{M_{\sun}}$ non-rotating stellar model, which provides
an educative example (see Appendix~\ref{app:example_models} for further examples). 
As explained above, $\Gamma'$ and $\Gamma_\deb{e}$ are significantly greater than one 
in the convective core of the star. The indicated sub-surface convection zones are caused 
by the opacity peaks at $T\sim 1.5\times 10^6$\, K (deep iron bump) and at 
$T\sim 2\times 10^5$\, K (iron bump). 
Near the bottom of the inflated envelope ($r/r_{\deb{core}}\gtrsim 1$; 
see Sect.~\ref{sec:inflation} for the definition of $r_{\deb{core}}$), $\Gamma$ 
approaches one and the Fe opacity bump 
drives convection. An extended region with $\Gamma \approx 1$ follows. 
A thin shell very close to the photosphere 
contains the layers with a positive density gradient 
and with $\Gamma >1$.

\begin{figure}[h!]
\centering
\resizebox{\hsize}{!}{\includegraphics[width=\linewidth, angle=-90]{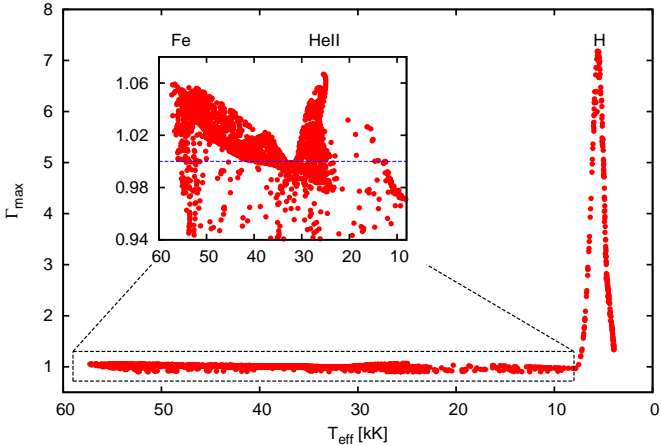}}
   \caption[]{Maximum Eddington factor in the analysed stellar models 
     as a function of $\mathrm{T_{eff}}$ for all stellar models 
     shown in Fig.~\ref{fig:gamma_HR}, i.e. with $\Gamma_{\mathrm{max}} >0.9$. 
     The three peaks in $\Gamma_{\mathrm{max}}$ at $\mathrm{T_{eff}/kK}$ of $\sim$ 55, 25 and 5.5 
     correspond to the three opacity bumps associated with the ionization zones of Fe, HeII and H respectively. 
     \textit{Inset}: Models with $\Gamma_{\mathrm{max}} >0.94$ and 
     $\mathrm{T_{eff}/kK}\gtrsim 10\,$kK. The blue horizontal 
     line at $\Gamma_{\mathrm{max}}=1$ is drawn for reference.} 
     \label{fig:gamma_Teff}
\end{figure}

\begin{figure}[h!]
\centering
\resizebox{\columnwidth}{!}{\includegraphics[angle=-90]{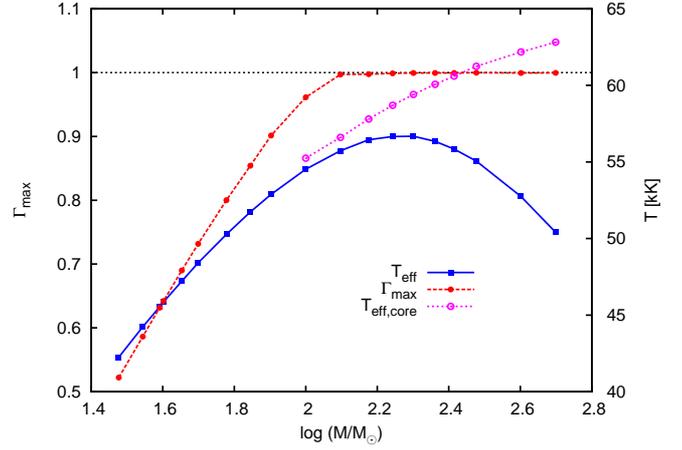}}
   \caption[]{Maximum Eddington factor ($\Gamma_{\mathrm{max}}$, red line), 
   effective temperature ($T_{\mathrm{eff}}$, blue line) and effective temperature 
   at the non-inflated core ($T_{\mathrm{eff,core}}$, pink line, 
   see Sect.\,\ref{sec:inflation}, eqn.\,\ref{eq:teff_core}) as a function of initial mass 
   for the non-rotating ZAMS models. The black dotted line at $\Gamma_{\mathrm{max}}=1$ is drawn for reference.}
\label{fig:gamma_ZAMS}
\end{figure}

\begin{figure*}[]
\centering
 \resizebox{\textwidth}{!}{\input{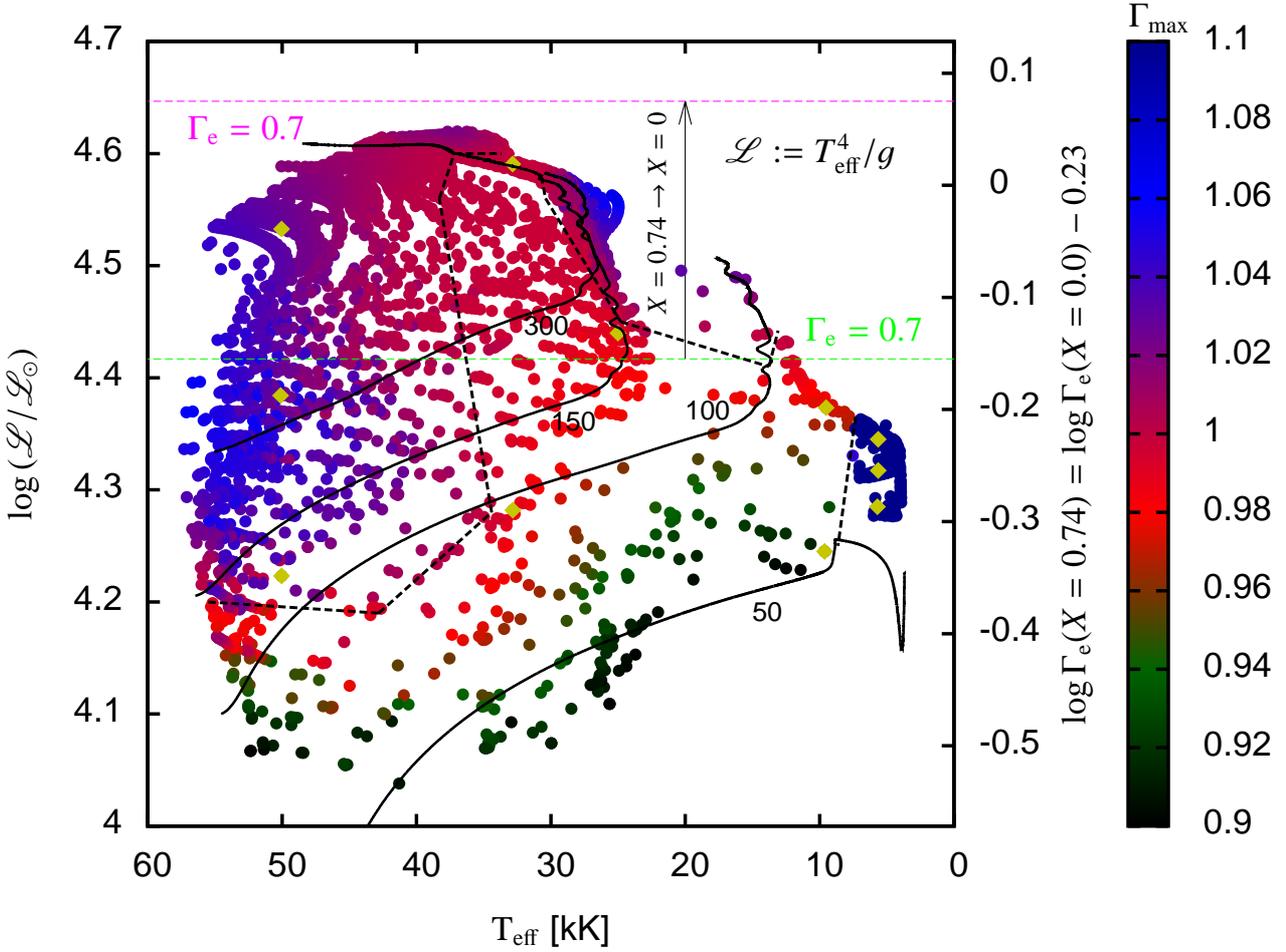}}
    \caption[]{Analyzed models shown as colored dots on the spectroscopic HR diagram (sHRD) with 
    the color representing the value of $\Gamma_{\deb{max}}$ in each model.  
    The Eddington factor $\Gamma_{\mathrm{e}}$
    (assuming electron scattering opacity only) is directly proportional 
    to the quantity $\mathscr{L}$ and is indicated on the 
    right Y-axis for a hydrogen mass fraction of $X=0.74$. 
    Some representative evolutionary tracks of non-rotating models, for different initial masses (indicated along the tracks in units of solar mass), are also shown with solid black lines.
    The green and blue dotted horizontal 
    lines correspond to $\Gamma_{\mathrm{e}}=0.7$ for $X=0.7$ and $X=0$ respectively.
    The color palette and the models marked with yellow diamonds 
    correspond to those in Fig.~\ref{fig:gamma_HR}. The black dashed line roughly 
    divides the sHRD into distinct regions with $\Gamma_{\deb{max}}>1$ and 
    $\Gamma_{\deb{max}}<1$. 
} 
\label{fig:gamma_sHRD}
\end{figure*}

\begin{figure}
\centering
 \resizebox{\hsize}{!}{\includegraphics[angle=-90]{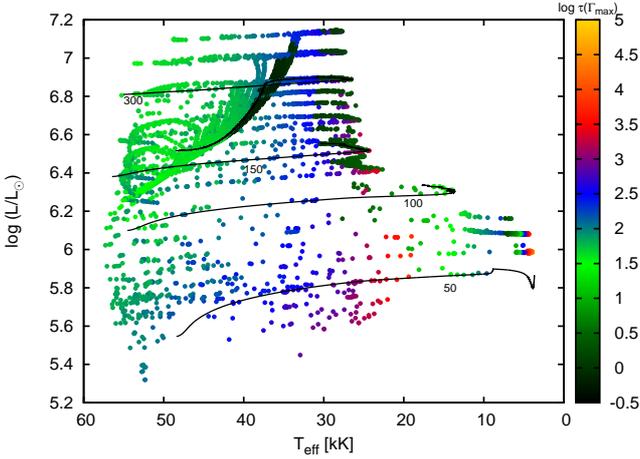}}
   \caption[]{Hertzsprung-Russell diagram showing the logarithm of the 
   optical depth $\tau$ at the position of $\Gamma_{\mathrm{max}}$ in color, for all analysed 
   models which have $\Gamma_{\deb{max}}>0.9$. 
       Some representative evolutionary tracks of non-rotating models, for different initial masses (indicated along the tracks in units of solar mass), are also shown with solid black lines.
       } 
\label{fig:gamma_tau}
\end{figure}

\begin{figure}[h!]
\centering
 \resizebox{\columnwidth}{!}{\input{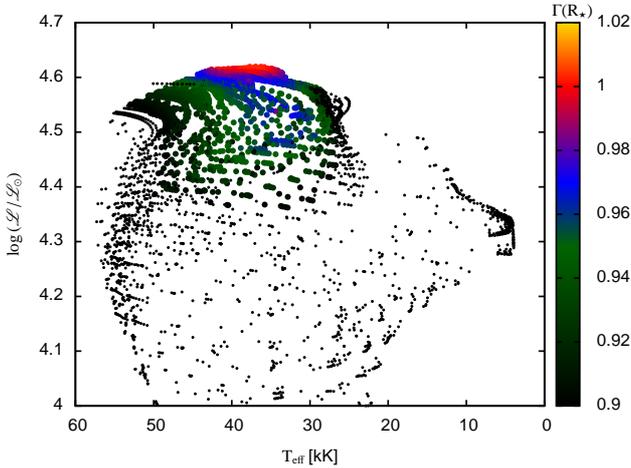}}
   \caption[]{The Eddington factors at the stellar surface, 
   $\Gamma \mathrm{(R_{\star})}$, are 
   shown on the sHRD for our analysed set of models. The models with 
   $\Gamma \mathrm{(R_{\star})}<0.9$ are indicated with smaller black circles and those 
   with $\Gamma \mathrm{(R_{\star})}>0.9$ are shown as colored dots.}
   \label{fig:surface_gamma_sHRD}
\end{figure}

We note that the stellar models have been computed with a hydrodynamic stellar evolution
code. However, due to the large time steps required for stellar evolution calculations,
non-hydrostatic solutions are suppressed by our numerical scheme. The resulting
hydrostatic structures are still valid solutions of the hydrodynamic equations
(see \citet{heger2000} and \citet{kozyreva_2014} for the equations
to be solved). Models computed with time steps small enough to resolve the hydrodynamic
time scale reveal that some, and potentially many, of our models are pulsationally unstable,
as will be shown in a forthcoming paper.
However, in the cases analysed so far, the pulsations saturate and do not lead to
a destruction or ejection of the inflated envelopes. In this respect, we consider
the analysis of the hydrostatic equilibrium structures performed in this paper
as useful. 

\subsection{Effect of rotation on the Eddington limit}
The effect of the centrifugal force on the structure of rotating stellar models has been 
studied by a number of groups in the past, including \citet{heger2000} and \citet{mm_2000}.
This is done by describing the models in a 1-D approximation where all variables are
taken as averages over isobaric surfaces \citep{kipp_thomas_1970}. 
The stellar structure equations are modified to include the effect of the centrifugal force
\citep{endal_1976}. The equation of hydrostatic equilibrium becomes
\begin{equation}
 \frac{dP}{dm}4\pi r^2 + f_{P} \frac{GM(r)}{r^2} =  0,  
 \end{equation}
 and the radiative temperature gradient in the 
 energy transport equation (in the absence of convection) takes the form 
 \begin{equation}
  \nabla_{\deb{rad}} = \frac{3}{16\pi acG}\frac{\kappa PL}{MT^4}\frac{f_T}{f_P},
 \end{equation}
 where the quantities $f_P$ and $f_T$ have the same definition 
 as in \citet{heger2000}.
Consequently, the Eddington luminosity gets modified as: 
 \begin{equation}
    L_{\deb{Edd}} = \frac{4\pi cGM}{\kappa} \frac{f_P}{f_T} 
 \end{equation}
However, the Eddington factor, 
\begin{equation}
 \Gamma = \frac{L_{\deb{rad}}}{L_{\deb{Edd}}}= \frac{\nabla}{\nabla_{\deb{rad}}}\frac{L}{L_{\deb{Edd}}} = \frac{4a}{3}\frac{T^4 \nabla}{P}
\end{equation}
does not have any explicit dependence on $f_P$ and $f_T$ because the factor $f_P/f_T$ cancels out.
Therefore formally, the Eddington factor remains unaffected by rotation. Of course,
if the internal evolution of a rotating model is changed, 
for example by rotational mixing,
its Eddington factor will still be different from 
that of the corresponding non-rotating model.  

Of course, real stars are three-dimensional and the centrifugal force must affect the hydrostatic 
stability limit. However, this is expected to be a function of the latitude at the stellar surface,
and in a 2-D view, the effect will be largest at the equator \citep{langer97}. To first order, 
the critical luminosity $L_{\rm c}$ to unbind matter at the stellar equatorial surface becomes
\begin{equation}
L_{\rm c} = L_{\rm Edd} \left( 1- \left(\frac{v_{\rm rot}}{v_{\rm Kep}}\right)^2 \right) ,
\end{equation}
where $v_{\rm rot}$  and $v_{\rm Kep}$ are the stellar equatorial rotation velocity and the
corresponding Keplerian value, respectively. However, to compute the effect reliably, the
stellar deformation due to rotation as well as the effect of gravity darkening need to be 
accounted for \citep{mm_2000,maeder_book_2009}. To do this realistically for stars near the Eddington limit requires
at least 2-D calculations. 

The implication is that the effect of rotation on the critical stellar luminosity
can not be properly described through the models analysed here. Those
models see the same critical luminosity as if rotation was absent. Since mixing of helium
in these models is very weak for rotation rates below the ones required for chemically 
homogeneous evolution, most of the rotating models evolve very similar to the
non-rotating ones \citep{ines2011,koehler2015}, and thus
merely serve to augment our database.

\subsection{The maximum Eddington factor}
In our stellar models, we have determined the 
maximum Eddington factor $\Gamma_{\mathrm{max}}$ over the whole star, i.e
$\Gamma_{\mathrm{max}} :=\max\, [\Gamma (r)]$. 
The maximum Eddington factor $\Gamma_{\mathrm{max}}$ generally
occurs in the outer envelopes of our models, where convective energy transport is 
much less efficient than in the deep interior. The variation 
of $\Gamma_{\mathrm{max}}$ across the upper HR diagram is shown 
in Fig.~\ref{fig:gamma_HR} for all analysed core hydrogen burning models 
which have $\Gamma_{\mathrm{max}} >0.9$. 

Three distinct regions with $\Gamma_{\mathrm{max}} >1$ can be 
identified in Fig.~\ref{fig:gamma_HR} which can be connected to 
the opacity peaks of iron, helium and hydrogen. 
When one of these opacity peaks is situated sufficiently close  
to the stellar photosphere, the densities in these layers are so small that 
convective energy transport becomes inefficient. As a consequence, 
super-Eddington layers develop which are stabilized by a
positive (i.e. inward directed) gradient in density and gas pressure   
(see Sect.~\ref{sec:dens_inv} below). 
The envelope inflation which occurs when $\Gamma_{\mathrm{max}}$ approaches one
is discussed in Sect.~\ref{sec:inflation}.

Figure \ref{fig:gamma_Teff} shows $\Gamma_{\mathrm{max}}$ 
as a function of the effective temperature 
for all our models with $\Gamma_{\mathrm{max}}>0.9$. 
The models which have the hydrogen opacity bump close to their 
photosphere can obtain values of $\Gamma_{\mathrm{max}}$ as high as $\sim 7$. 
This manifests itself as a prominent peak around $\mathrm{T_{eff}} \approx 5.5\,$kK. 
The inset panel shows the much weaker peaks in $\Gamma_{\mathrm{max}}$ due to the partial 
ionization zones of Fe and HeII, at $\mathrm{T/kK}\sim 200$ and 
$50$ respectively. The peak caused by the Fe opacity bump may extend to
hotter effective temperatures and apply to hot, 
hydrogen-free Wolf-Rayet stars, which are not
part of our model grid.

For stars above about $125\mso$, $\Gamma_{\mathrm{max}}$ 
reaches one, even on the zero-age main sequence.
This is demonstrated in Fig.~\ref{fig:gamma_ZAMS} which shows both
$\Gamma_{\mathrm{max}}$ and effective temperature
as a function of mass for the non-rotating stellar
models. As these models evolve away from the ZAMS 
to cooler temperatures, super-Eddington layers develop in their interior.  
The blue curve shows a maximum effective temperature of $57\,000$\,K at about $200\mso$,
beyond which it starts decreasing with further increase in mass.
This behaviour is related to the phenomenon of inflation which is discussed in detail
in Sect.~\ref{sec:inflation}. However, the `effective temperature' at the 
base of the extended, inflated envelope, $T_{\rm eff,core}$ 
(see Sect.\,\ref{sec:inflation}), still increases with 
mass in the whole considered mass range.

\subsection{The spectroscopic HR diagram}

Figure \ref{fig:gamma_sHRD} shows the location of our analysed models 
in the spectroscopic HR diagram (sHRD) \citep{langer2014, koehler2015} 
where instead of the 
luminosity, $\mathscr{L}:=T_{\mathrm{eff}}^4/g$ is plotted 
as a function of the effective temperature. The quantity $\mathscr L$ can be measured 
for stars without knowing their distance. 
Moreover, we have $\log (\mathscr L / \mathscr L_{\odot}) =  \mathcal{R} $ (cf., Sect.\,1), 
such that $\mathscr{L}$ is directly 
proportional to the Eddington factor $\Gamma_{\mathrm{e}}$ as
\begin{equation}
 \Gamma_{\mathrm{e}} = \frac{\kappa_{\deb e} L}{4\pi cGM}=\frac{\kappa_{\deb e} \sigma T_{\mathrm{eff}}^4}{cg}=\frac{\kappa_{\deb e} \sigma}{c}\mathscr {L},
\end{equation}
where  $g$ is the surface gravity 
and the constants have their usual meaning.
Therefore one can directly read off $\Gamma_{\mathrm{e}}$ 
(right Y-axis in Fig.~\ref{fig:gamma_sHRD}) from the sHRD. 
Massive stellar models often evolve with a slowly increasing
luminosity over their main-sequence lifetimes. 
Therefore, where in a conventional HR diagram models with very different 
$\Gamma_{\mathrm{max}}$ might cluster together (see Fig.~\ref{fig:gamma_HR}),
they separate out nicely in the sHRD since $\mathscr {L}\propto L/M$. 
The effect of the opacity peaks on the 
maximum Eddington factor ($\Gamma_{\mathrm{max}}$) at temperatures 
corresponding to the three partial ionization zones 
(Fe, HeII and H) is seen more clearly in
the sHRD in Fig.~\ref{fig:gamma_sHRD} compared to the ordinary HR diagram (Fig.~\ref{fig:gamma_HR}). 

We find that for our ZAMS models the electron scattering 
opacity is $\kappa_{\mathrm{e}} \approx 0.34$ while 
the true photospheric opacity $\kappa_{\mathrm{ph}}$ 
is around 0.5. Therefore it is expected that the true 
Eddington limit ($\Gamma=1$) 
is achieved at about $\Gamma_{\mathrm{e}}=0.7$ for stellar models which retain 
the initial hydrogen abundance at the photosphere. Therefore in Fig.~\ref{fig:gamma_sHRD} 
we have drawn two horizontal lines corresponding to $\Gamma_{\mathrm{e}}=0.7$, one assuming 
the initial hydrogen mass fraction $X=0.74$ (green line) 
and the other assuming $X=0$ (red line). 
While models with helium-enriched photospheres exceed the green line comfortably, even the 
most helium-enriched models (rotating or otherwise) stay below the red line.  

From \citet{koehler2015}, we know that
models with $\log \mathscr{L} / \mathscr{L}_{\odot} > 4.4$ are 
all hydrogen-deficient, either due to mass loss or due to rotationally induced mixing,
as both processes lead to an increasing $L/M$-ratio (cf., their fig.\,18).
Figure \ref{fig:gamma_sHRD} thus demonstrates that the models which contain
super-Eddington layers due to the partial ionization of helium all have
hydrogen deficient envelopes, i.e., they are correspondingly helium-enriched.   

Figure \ref{fig:gamma_sHRD} reveals that the electron scattering 
Eddington factor $\Gamma_{\deb e}$ is not a good proxy for the maximum 
true Eddington factor ($\Gamma_{\deb{max}}$) obtained inside the star. 
For example, along the horizontal line $\log {\mathscr L}/{\mathscr L}_{\sun}=4.3$, 
corresponding to $\Gamma_{\deb e}\simeq0.5$, $\Gamma_{\deb{max}}$ varies from 
well below one to values near seven at the cool end.
However, we note that below $30\,000$\,K helium and hydrogen recombine, and the gas is not fully ionised any more. 
The line opacities of helium and hydrogen become important, 
which causes the increase in $\Gamma_{\rm max}$ (see Figs.\,\ref{fig:gamma_Teff} and \ref{fig:opal}).

\begin{figure}[h!]
\centering
\resizebox{\columnwidth}{!}{\includegraphics[angle=-90]{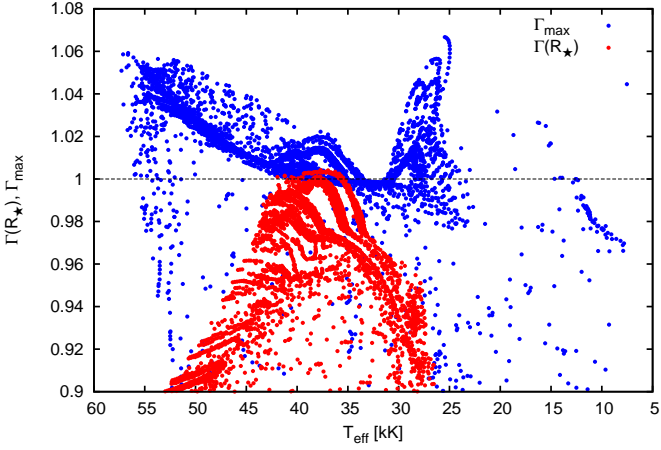}}
   \caption[]{Maximum Eddington factor ($\Gamma_{\mathrm{max}}>0.9$) and surface 
   Eddington factor ($\Gamma{\mathrm{(R_{\star})}}>0.9$) as a function of the 
   effective temperature $T_{\mathrm{eff}}$ of our analysed models. 
   The black dotted line at $\Gamma =1$ is drawn for convenience.}
\label{fig:gamma_surf_gammax}
\end{figure}

\subsection{Surface Eddington factors and the location of $\Gamma_{\deb{max}}$}

The optical depth where $\Gamma_{\deb{max}}$
is reached gives an idea of
how deep in the stellar interior the layer
with the highest Eddington factor is located.
We investigate this in Fig.~\ref{fig:gamma_tau} which shows that
$\Gamma_{\deb{max}}$ is
located at largely different optical depths in different
types of models.
While the maximum Eddington factors occur generally at optical depths below $\sim 10\,000$,
we see that in the three effective temperature regimes identified by the
super-Eddington peaks in Figs.\,\ref{fig:gamma_HR} and\,\ref{fig:gamma_Teff},
$\Gamma_{\deb{max}}$ can even be located at an optical depth of $\sim 10$ or below.

For example, 
when the tracks above $\log L/L_{\sun}=6.2$ approach 
effective temperatures of $\sim 30$\,kK, 
$\Gamma_{\deb{max}}$ is located at 
the Fe-peak which is deep inside the 
envelope ($\tau\simeq 1400$). 
The models at this stage 
become helium-rich ($Y_s\gtrsim 70\%$) and the
Wolf-Rayet mass loss prescription is applied. 
Once these tracks turn bluewards
in the HR diagram, the position of
$\Gamma_{\deb{max}}$ jumps to the helium opacity peak, which is located
much closer to the stellar surface. Consequently, we find
three orders of magnitude of difference between these two types of models
with similar effective temperature and luminosity.

When considering the surface Eddington-factors in the spectroscopic
Hertzsprung-Russell diagram (Fig.~\ref{fig:surface_gamma_sHRD}),
we see that only the models with $\Gamma(R_{\star}) > 0.98$ 
have $\log \mathscr{L}/\mathscr{L}_{\odot}$ values 
of more than $4.6$. As discussed above, these models,
which started on the main sequence with initial masses above $300\mso$   
are extremely helium-rich and may correspond to the most extreme
late-type WNL stars \citep{sander_2014}. 
As shown in Fig.~\ref{fig:gamma_surf_gammax},
they exceed the Eddington limit by just a few per mill, which is possible
because of the high assumed mass loss rates that 
imply a slight deviation from hydrostatic equilibrium near 
the stellar surface (cf. Sect.~\ref{sec:edd_limit} above). 
However, these models are the ones where our assumption of an optically
thin wind might break down \citep[see Fig.\,7 in][]{koehler2015}. 
Since the inclusion of an optically thick outflow
may lead to changes of the temperature and density structure near the surface, 
the surface Eddington-factors for these particular models are not reliable.

In summary, we find on one hand that many of our 
models contain layers at optical depths between a few and 
a few thousand in which the Eddington factor exceeds the critical value of one.
On the other hand, for none of our models we can conclude that the Eddington limit 
is reached very near to, or at the surface, where for the vast majority we can even
exclude that this happens. This finding leads to a shift in the expectation of
the response of stars that reach the Eddington limit during their evolution.
We might not expect direct outflows driven by super-Eddington luminosities,
but instead internal structural changes, in particular envelope inflation.


\section{Envelope inflation}\label{sec:inflation}
\begin{figure}
\centering
\resizebox{\columnwidth}{!}{\includegraphics[angle=-90]{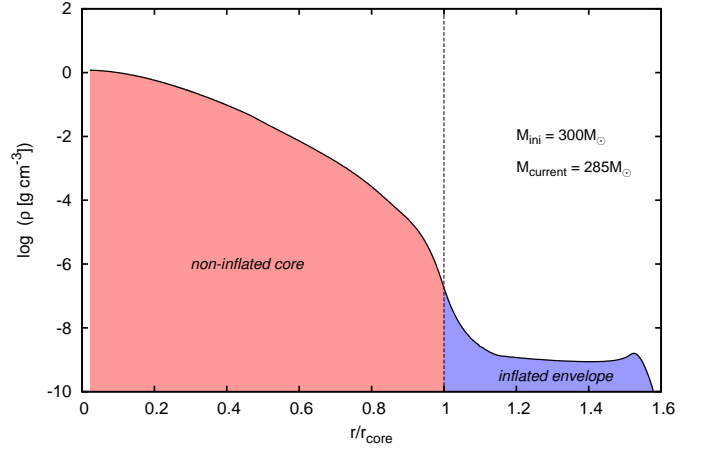}}
   \caption[]{
     Density profile of a non-rotating $285\,\mathrm{M_{\sun}}$ star with 
     $T_{\mathrm{eff}}=46600\,$K and $\log (L/L_{\odot})=6.8$ (cf. Fig.\,\ref{fig:gamma_eg} 
     and Appendix \ref{app:inflations_eg}) showing 
     an inflated envelope and a density inversion. 
     The X-axis has been scaled with the core radius $r_{\mathrm{core}}$ of $25.3 \rso$, as
     defined in Sect.\, \ref{sec:inflation}.} 
\label{fig:inflation_eg}
\end{figure}

\emph{Inflation} of massive, luminous stars refers to the formation of extended, 
extremely dilute stellar envelopes.
An example of an inflated model is shown in Fig.~\ref{fig:inflation_eg}. The 
red shaded region is the non-inflated \textit{core} and the blue shaded region is what 
we refer to as the inflated envelope. In the example, the model is inflated by 60\% 
of its core radius (defined below). In the presented model, 
the inflated envelope only contain a small fraction
of a solar mass, i.e. $\approx 10^{-5}\,\mso$.

Envelope inflation is inherently different from classical red supergiant formation.  
The latter occurs after core hydrogen exhaustion, as a consequence of vigorous
hydrogen shell burning. This process expands all layers above the shell source, which
usually comprise several solar masses in massive stars, and it also operates in low
mass stars, such that no proximity to the Eddington limit is required.
The mechanism of envelope inflation which we discuss here works 
already during core hydrogen burning, i.e., even on the zero-age 
main sequence for sufficiently luminous stars (cf., Fig.\,\ref{fig:gamma_ZAMS}).
Previous investigations have suggested that inflation is related to the
proximity of the stellar luminosity to the Eddington luminosity
\citep{ishii99,petrovic_2006,graefener_2012} in the envelopes of massive stars 
with a high luminosity-to-mass ratio ($\gtrsim 10^{4}\, \mathrm{L_{\sun}}/\mathrm{M_{\sun}}$).
The amount of mass contained in an inflated envelope is usually 
very small. As we shall see below, inflation, in extreme cases, can also produce
core hydrogen burning red supergiants. 

 \begin{figure}
\centering
 \resizebox{\hsize}{!}{\includegraphics[angle=-90]{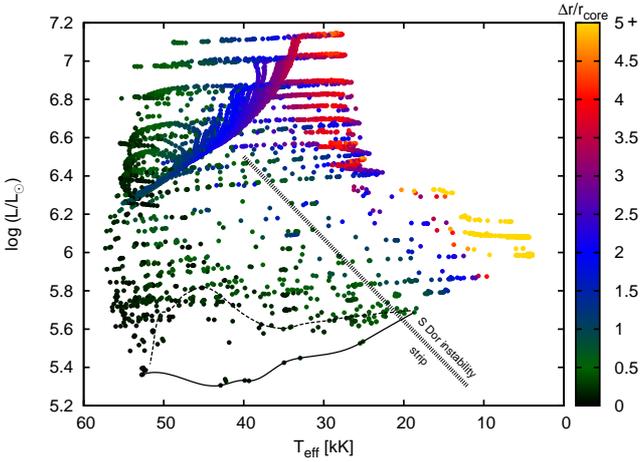}}
   \caption[]{Hertzprung-Russell diagram showing all the core-hydrogen burning inflated models, 
   i.e with $\Delta r/r_{\deb{core}}>0$. Models with $\Delta r/r_{\deb{core}}>5$ 
   are colored yellow. Below the solid black line we do not find any inflated models and  
 above the dotted black line we do not find any non-inflated models in our grid. The hot 
 part of the S\,Dor instability strip is also marked \citep{smith2004}.}
\label{fig:inflation_grid}
\end{figure}

We define inflation in our models through
$\Delta r/r_{\deb{core}}:=(R_{\star}-r_{\mathrm{core}})/r_{\mathrm{core}}$, 
with $r_{\mathrm{core}}$ being the radius at which 
inflation starts and $R_{\star}$, the photospheric radius. 
Since the densities in inflated envelopes are small, the dominance of radiation pressure
in these envelopes is much larger than it is in the main stellar body.
We define a model to be inflated if $\beta(r)$, which is the ratio of 
gas pressure to total pressure, reaches a value below 0.15 in the interior of a model. 
The radius at which $\beta$ goes below 0.15 for the first time from the center outwards 
is denoted as $r_{\mathrm{core}}$, i.e. the start of the inflated region. The remaining extent 
of the star until the photosphere $(R_{\star}-r_{\mathrm{core}})$ is considered 
as the inflated envelope.  

We emphasize that our choice of the threshold 
value for $\beta$ is arbitrary and not derivable from first principles. However, 
we have verified that this prescription identifies inflated stars in different parts 
of the HR diagram very well (cf. Appendix \ref{app:example_models}). 
As $\beta \rightarrow 0$ for $M \rightarrow \infty$, our criterion may fail for extreme
masses. However, the mass averaged value of $\beta$ for the most massive zero-age main sequence model 
analysed in the present study ($500\,\deb{M_\sun}$) is $0.3$. A threshold value of $0.15$ thus appears
adequate for the present study. As an example, 
let us consider a typical inflated model, shown in Appendix \ref{app:inflations_eg}.
The value of $\beta$ in Fig.\,\ref{fig:app:beta} decreases 
sharply at the base of the inflated envelope, 
to around $0.01$. Even if the $\beta$ threshold is varied by 30\%, i.e. $0.15\pm 0.045$, 
the non-inflated core radius $r_{\deb{core}}$ changes by only 4\%. 
This goes to show that for clearly inflated models, 
the value of $r_{\deb{core}}$ is insensitive to the threshold value of $\beta$. 

We furthermore performed a numerical experiment which is suited to show
that the core radii identified as described above are indeed robust. 
We chose an inflated $300\,\mso$ model, and then increased the 
mixing length parameter $\alpha$ such that convection 
becomes more and more efficient. As shown in Fig.\,\ref{fig:app:mix_length_comparison},
as a result the extent of the inflated envelope decreased without
affecting the model structure inside the core radius, which thus
remained independent of $\alpha$. For $\alpha=40$ convection became nearly adiabatic,
inflation almost disappeared, and the fact that the photospheric radius in this case
became very close to the core radius 
validates our method of identifying $r_{\rm core}$.

Figure \ref{fig:inflation_grid} shows the amount of inflation as defined above, for all
our models that fulfil the inflation criterion, in the HR diagram. 
It reveals that overall, inflation is larger for cooler temperatures. This
is not surprising, since inflation appears not to change the stellar luminosity 
and must therefore induce smaller surface temperatures.
We also see that inflation is larger for more luminous stars, which
is expected because the Eddington limit is supposed to play a role (see below). 
We also find inflation to increase along 
the evolutionary tracks of the most massive stars 
which turn back from the blue supergiant stage, 
which in this case is due to the shrinking of their core radii. 
A distinction between 
the inflated and the non-inflated models 
is made by drawing the black lines in Fig.~\ref{fig:inflation_grid}. 
They are drawn such that below the solid line no model is inflated  
and above the dotted line all the models are inflated. In between these 
two lines we find a mixture of both inflated and non-inflated models. 
We find that essentially all models above $\log (L/\lso) \simeq 5.6$ are 
inflated. Consequently, stars above $\sim 40 \mso$ do inflate during 
their main sequence evolution.

 \begin{figure}
\centering
 \resizebox{\hsize}{!}{\includegraphics[angle=-90]{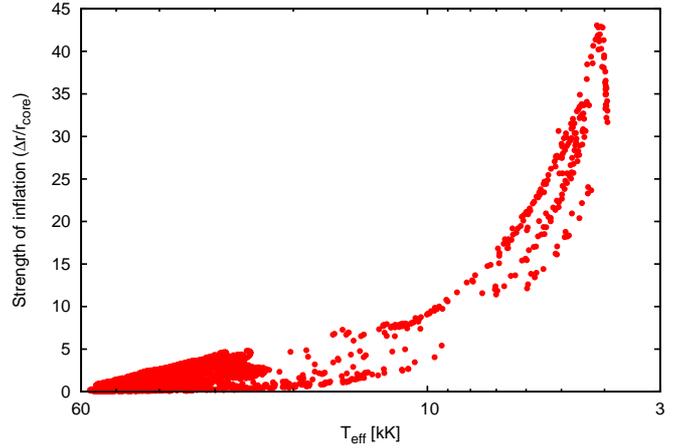}}
   \caption[]{Inflation as a function of the effective temperature 
   for all analysed models which fulfill our inflation criterion.}
\label{fig:inflation_Teff}
\end{figure}

Figure \ref{fig:inflation_Teff} shows the inflation factor as function of the stellar
effective temperature for our inflated models.
Whereas inflation increases the radius of our hot stars by up to a factor of 5,
the cool supergiant models can be inflated by a factor of up to 40. 
We refer to Appendix \ref{app:inflations_eg} for the detailed structures 
of several inflated models. 

\begin{figure}
\centering
\resizebox{\hsize}{!}{\includegraphics[angle=-90]{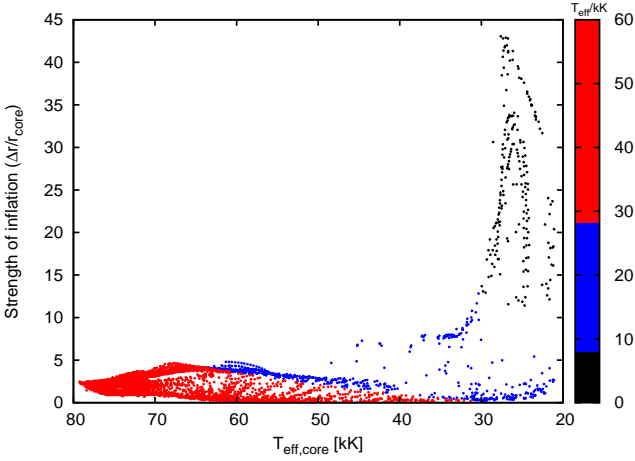}}
   \caption[]{Inflation as a function of the effective temperature at the 
   core-envelope boundary for all analysed models with $\Delta r/r_{\deb{core}}>0$. Color coding indicates the 
   $T_{\mathrm{eff}}$ at the photosphere.} 
\label{fig:inflation_Teff_core}
\end{figure}

In Fig.~\ref{fig:inflation_Teff_core},
we take a look at inflation as a function of the core effective temperature 
$T_{\mathrm{eff,core}}$ defined as 
\begin{equation}\label{eq:teff_core}
T_{\mathrm{eff,core}}=\frac{L}{4\pi \sigma r_{\mathrm{core}}^2},
\end{equation}
where $L$ refers to the surface luminosity and 
the constants have their usual meaning.  
We can see that even our coolest models have 
high core effective temperatures, in the sense that
if their inflated envelopes were absent, 
their stellar effective temperatures would 
have been higher than 20\,000\,K. Those stars 
which have stellar effective temperatures below 
$\sim$50\,000\,K contain the He\,II 
ionization zone within their envelopes, and
stars with stellar effective temperatures 
below $\sim$10\,000\,K also contain the
H/He\,I ionization zone. However, as 
revealed by the density and temperature structure
of these models (cf., Appendix \ref{app:example_models}), 
the temperature at the bottom of the inflated envelope
is always about 170\,000\,K, and thus corresponds to the 
temperature of the iron opacity peak.
We conclude that the iron opacity is at least in 
part driving the inflation of all the stars.
For those with cool enough envelopes, helium and 
hydrogen are likely relevant in addition.

 \begin{figure}
\centering
\resizebox{\hsize}{!}{\includegraphics[angle=-90]{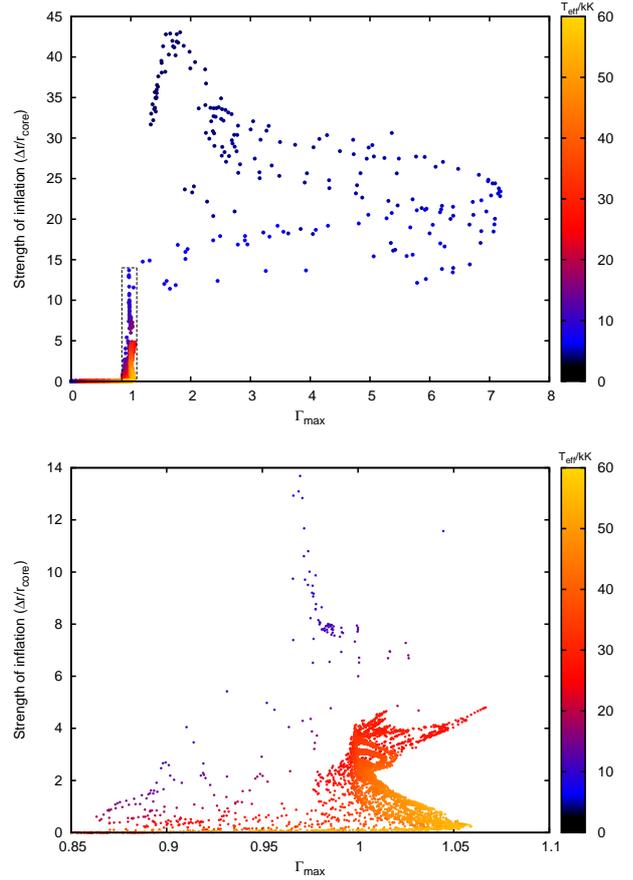}}
   \caption[]{\textit{Top:} Inflation ($\Delta r/r_{\deb{core}}$) as a function of $\Gamma_{\mathrm{max}}$ for the analysed models.  
   the effective temperature of every model is color-coded.
   The area within the black dotted lines is magnified below.
   \textit{Bottom:} Zoomed-in view of the dotted region in the top panel around $\Gamma_{\deb{max}}=1$. }
\label{fig:inflation_gamma}
\end{figure}

\subsection{Why do stellar envelopes inflate?}\label{subsec:discussion}

As suggested earlier, the physical cause of inflation in a given star may be its proximity 
to the Eddington limit.  Figure \ref{fig:inflation_gamma} shows the correlation between 
inflation and $\Gamma_{\mathrm{max}}$ for our models.
As expected, we find that our stellar models are not inflated when $\Gamma_{\mathrm{max}}$ 
is significantly below 1, and they are all inflated for $\Gamma_{\mathrm{max}} > 1$.
Indeed, the top panel of Fig.~\ref{fig:inflation_gamma} gives the clear message
that the Eddington limit, in the way it is defined in Sect.\,\ref{sec:edd_limit}, is likely connected with envelope 
inflation.

Comparing Fig.~\ref{fig:inflation_gamma} (top panel) to Fig.~\ref{fig:inflation_Teff}
shows that inflation 
increases up to $T_{\mathrm{eff}}\approx 5\,500\,$K.  
Thereafter, $T_{\mathrm{eff}}$ and $\Gamma_{\mathrm{max}}$ decrease and  
the stars keep getting bigger without significant changes in $r_{\mathrm{core}}$,
and hence, inflation still increases. 
However, the drop in inflation for the 
coolest models shows an opposite trend. This is because the non-inflated core radius 
$r_{\mathrm{core}}$ now moves outwards (increases) 
such that inflation ($\Delta r/r_{\deb{core}}$) decreases even though  
$R_{\star}$ keeps increasing (cf. definition of $r_{\deb{core}}$ in Sec.\,\ref{sec:inflation}).

In the zoom-in at the lower panel of Fig.~\ref{fig:inflation_gamma}, 
we see some models being inflated for 
$\Gamma_{\mathrm{max}}$ in the range $\sim 0.9\dots 1$. 
Partly, this may be due to the arbitrariness in our definition of inflation.
The exact value of $\Delta r/r_{\deb{core}}$ depends somewhat on the choice 
of the threshold value of $\beta$ to characterize inflation 
(cf. Appendix \ref{app:example_models}), 
i.e., the models with $\Delta r/r_{\deb{core}} \lesssim 2$ 
and $\Gamma_{\mathrm{max}} < 1$ may be at
the borderline of inflation. The models with $\Delta r/r_{\deb{core}} \lesssim 2$ 
but $\Gamma_{\mathrm{max}} > 1$ are all very hot ($T_{\mathrm{eff}} \simgr 40\,000\,$K) and 
in those models, the inflation is intrinsically small, but generally unambiguous. 

Still, we see a significant number of models below the Eddington limit ($\Gamma_{\mathrm{max}} < 1$)
which show a quite prominent inflation, i.e. which have a radius increase due to inflation of more than
a factor of five. We investigated such a model by artificially
increasing its mass loss rate above the critical value $\dot{M}_{\rm crit}$ \citep{petrovic_2006}, such that  
the inflated envelope was removed (cf. Sect.\,\ref{subsec:masloss_inflation}).
We then found that, on turning down the mass loss rate to its original value, the model
regained its initial inflated structure with $\Gamma_{\mathrm{max}}<1$.
However, $\Gamma_{\mathrm{max}} = 1$  was reached and exceeded in the course
of our experiment. We conclude that a stellar envelope may remain inflated even if
the condition $\Gamma_{\mathrm{max}} = 1$ is not met any more in the 
course of evolution, but that $\Gamma_{\mathrm{max}} \gtrsim 1$ may be 
required to obtain inflation in the first place.

 \begin{figure}
\centering
\resizebox{\hsize}{!}{\includegraphics[angle=-90]{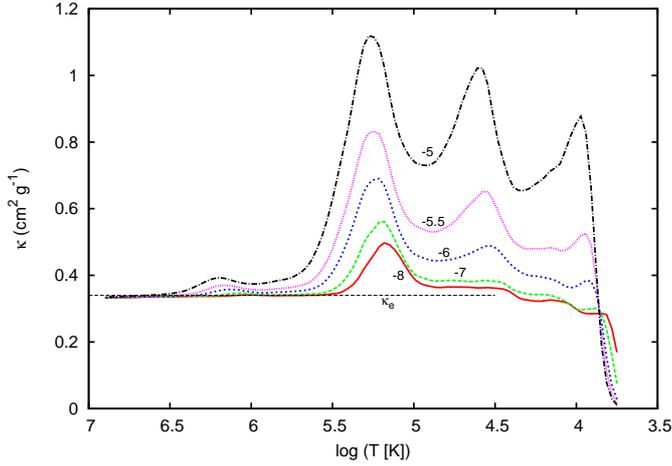}}
   \caption[]{Opacity as a function of temperature for fixed values 
   of the opacity parameter $R$ defined as $R=\rho/T_6^3$ where $T_6$ 
   is the temperature in units of $10^6$\,Kelvin. The values of $\log R$ which 
   are held constant are indicated along the curves. The data is taken 
   from the OPAL tables with a composition of X=0.7000, Y=0.2960, Z=0.004 \citep{ir96}. 
   The black horizontal line shows the electron scattering opacity $\kappa_\deb{e}$ 
   for a hydrogen mass fraction of $X=0.7$.
   At a temperature of 200\,000\,K, $\log R=-5$ implies a density of 
   $\rho = 8\times 10^{-8}\,{\rm g\,cm^{-3}}$ and $\log R=-8$ implies a density of $\rho = 8\times 10^{-11}\,{\rm g\,cm^{-3}}$ .}
\label{fig:opal}
\end{figure}

We see that in contrast to earlier ideas of a hydrodynamic outflow 
being triggered when the stellar surface  
reaches the Eddington limit \citep{eddington_1926,owocki2004}, 
in our models this never happens, but when
the properly defined Eddington limit is reached inside the envelope,
its outermost layers expands hydrostatically and produce inflation.
Two possibilities arise in this process. When the star 
approaches the Eddington limit, the ensuing
envelope expansion leads to changes in the temperature 
and the density structure. Consequently, the envelope
opacity can either increase or decrease. 
Fig.\,\ref{fig:opal} shows that the effect of 
expansion generally leads to a reduced opacity 
such that the expansion is indeed 
alleviating the problem. The star will then expand until
the Eddington limit is just not exceeded any more, which 
is the reason why we find so many inflated models 
with $\Gamma_{\mathrm{max}} \simeq 1$.  

Figure~\ref{fig:opal} shows the OPAL opacities for hydrogen-rich composition
for various constant values of $R$ as function of temperature, where 
$R=\rho/(T/10^6)^3$. \citet{kw90} showed 
that for constant $\beta = P_{\rm gas}/P$ and constant chemical
composition, $R$ as a function of spatial co-ordinate inside the star is a constant. 
Thus, for un-inflated models, the opacity curves in Fig. \ref{fig:opal}
may closely represent the true run of opacity with temperature
inside the star. In the inflated models, $\beta$ is dropping
abruptly at the base of the inflated envelope, which means that
the opacity is jumping from a curve with a higher $R$-value to one with a lower
$R$-value at this location. That is, the opacity is smaller everywhere
in the inflated envelope compared to the situation where inflation
would not have happened. 

For the chemical composition given in Fig.\,\ref{fig:opal} and assuming
$\beta \equiv\,$const., we find
\begin{equation}
R \simeq 1.8\, 10^{-5}  {\beta \over 1 - \beta} ,
\end{equation} 
such that if $\beta$ drops from 0.5 in the bulk of the star to 0.1 in
the inflated envelope, $R$ drops by one order of magnitude. 
The corresponding reduction in opacity can be significant, i.e., up
to about a factor of two.

When upon expansion the envelope becomes cool enough 
for another opacity bump to come into play,
the problem of not exceeding the Eddington limit might not be
solvable this way. 
Instead, when a new opacity peak is encountered in the outer part of the envelope, 
super-Eddington conditions occur, i.e.,
layers with  $\Gamma_{\mathrm{max}} > 1$ (cf., Figs.\,\ref{fig:gamma_HR} 
and \ref{fig:gamma_sHRD}), along with a 
strong positive gas pressure (and density) gradient 
(cf. Sects.\, \ref{sec:edd_limit} and \ref{sec:dens_inv}).
This is most extreme when the envelopes become 
cool enough ($T_{\mathrm{eff}} \simle 8000\,$K) such that 
the hydrogen ionization zone is present in the 
outer part of the envelope, where Eddington factors 
of up to seven are achieved.

\subsection{Influence of mass loss on inflation}\label{subsec:masloss_inflation}
One might wonder about the sustainability of the 
inflated layers against mass loss which is 
an important factor in the evolution of metal-rich massive stars. 
\citet{petrovic_2006} estimated that the 
inflated envelope can not be replenished
when the mass loss rate exceeds a critical value of
\begin{equation}\label{eqn:critical_massloss}
 \dot{M}_\mathrm{{crit}}=4\pi r_{\mathrm{core}}^2 \, \rho_{\mathrm{min}}\sqrt{\frac{GM}{r_{\mathrm{core}}}}\, ,
\end{equation}
where $M$ and $r_{\mathrm{core}}$ stand for the 
stellar mass and the un-inflated radius 
respectively, and $\rho_{\mathrm{min}}$ 
is the minimum density in the inflated region. 
\citet{petrovic_2006} found 
$\dot{M}_{\mathrm{crit}}\sim 10^{-5}\, \mathrm{M_{\sun}}\,{\rm yr}^{-1}$ 
for a massive hydrogen-free Wolf-Rayet star of $24\,\mathrm{M_{\sun}}$.
However, for a typical inflated massive star on the main sequence 
(see Fig.~\ref{fig:app:dens}), this critical mass loss rate is of the 
order $10^{-3}\ldots10^{-1}\, \mathrm{M_{\sun}}\,{\rm yr}^{-1}$. 
Such high mass loss rates are expected 
only in LBV-type giant eruptions. The mass loss rates applied to our models
are several orders of magnitude smaller \citep[cf.][]{koehler2015}.

\begin{figure}
\centering
\resizebox{\hsize}{!}{\includegraphics[angle=-90]{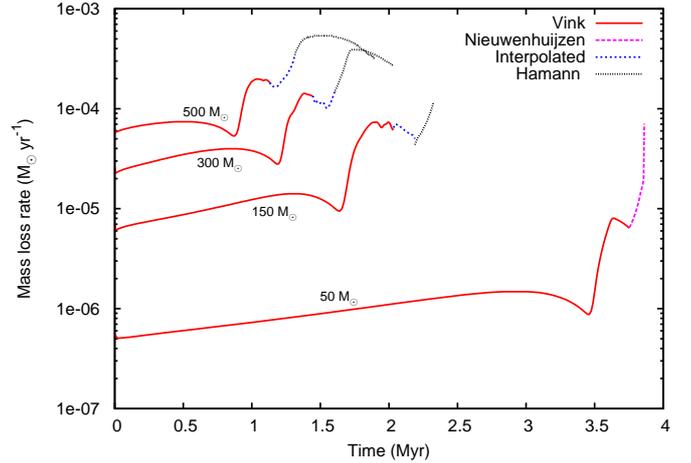}}
   \caption[]{Mass loss history of four non-rotating evolutionary sequences from our grid. The 
   initial masses are  
   given along each evolutionary track. The colours indicate the different mass loss prescriptions that 
   were used in different phases, as described in Sect.\,2.} 
\label{fig:massloss_history}
\end{figure}

The mass loss history of four evolutionary sequences without rotation are shown in Fig.\,\ref{fig:massloss_history}. 
We can see that even the $500\mso$ model never exceeds a mass loss rate  
$\sim 5\times 10^{-4}\,\mathrm{M_{\sun}}\,{\rm yr}^{-1}$. The critical 
mass loss rate for all models shown in Fig.\,\ref{fig:massloss_history} is 
much higher than the actual mass loss rates applied. Whereas $\dot{M}_\mathrm{{crit}}$ typically 
exceeds $\dot{M}$ by a factor of $1000$ for the inflated models in the $50\mso $ sequence, 
it exceeds that of the $500\mso$ sequence by a factor of $3\dots 100$.
It is thus not expected that mass loss prevents the formation of the inflated envelopes
in massive stars near the Eddington limit. In fact, it may be difficult to identify
a source of momentum that might drive such strong mass loss  \citep{shaviv2001,owocki2004}.
\citet{graefener2011} in the Milky Way and \citet{bestenlehner2014} 
in the LMC found a steep dependence of the
mass loss rates on the electron-scattering Eddington factor $\Gamma_\mathrm{{e}}$ for
very massive stars, but they do not find mass loss rates that 
substantially exceed $10^{-4}\,\mathrm{M_{\sun}}\,{\rm yr}^{-1}$.

As many of the models analysed here may be 
pulsationally unstable, the mass loss rates may be enhanced in this case.
\citet{grott2005} show that hot stars near 
the Eddington limit may undergo mass loss due to
pulsations, although extreme mass loss rates are not predicted. 
For very massive cool stars on the other hand,
\citet{moriya_langer_2015} find that 
pulsations may enhance the mass loss rate to values of
the order of $10^{-2}\, \mathrm{M_{\sun}}\,{\rm yr}^{-1}$.
Such extreme values could 
prevent the corresponding stars to spend 
a long time on the cool side of the Humphreys-Davidson limit.
A detailed consideration of this issue 
is beyond the scope of the present paper.

\section{Density inversions}\label{sec:dens_inv}

An inflated envelope can be associated 
with a `density inversion' near the stellar surface,
i.e., a region where the density increases outward. 
An example is shown in Fig.~\ref{fig:inflation_eg}.
In hydrostatic equilibrium, 
$\Gamma(r) >1$ implies $\frac{dP_{\mathrm{gas}}}{dr}>0$, and thus $\frac{d\rho}{dr}>0$. 
As a consequence, all the models which have layers 
in their envelopes exceeding the Eddington limit
show density inversions.
The criterion for density inversion can be expressed as \citep{joss73,mesa2013}:
\begin{equation}
\frac{L_{\mathrm{rad}}}{L_{\mathrm{Edd}}}>\left[ 1 + \left(\frac{\partial P_{\mathrm{gas}}}{\partial P_{\mathrm{rad}}}\right)_{\rho} \right]^{-1},
\end{equation}
where  $P_{\mathrm{gas}}$,  $P_{\mathrm{rad}}$ and $\rho$ stand for the gas pressure, 
radiation pressure and density respectively. A density inversion 
gives an inward force and acts as a stabilizing agent for the inflated envelope. We note that in the above 
inequality, $P_{\mathrm{gas}}$ is assumed to be a function of $\rho$ and $T$ only, i.e. the 
mean molecular weight $\mu$ is assumed to be constant. Density inversions might also be present 
in low-mass stars like the Sun where they are caused by the steep increase of $\mu$ 
around the hydrogen recombination zone \citep[cf.][]{ergma1971}.

 \begin{figure}
\begin{center}
\resizebox{\hsize}{!}{\mbox{\includegraphics[width=\linewidth, angle=-90]{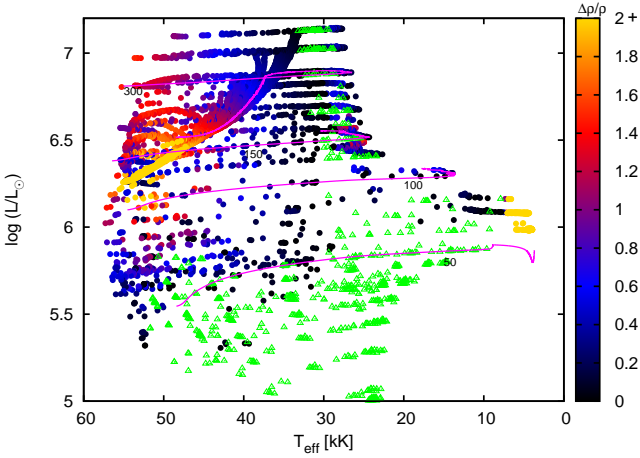}}}
   \caption[]{Upper HR diagram showing all main sequence models with density inversions. 
   Models with ${\Delta\rho}/\rho>2$ have been colored yellow. The models without density inversion 
   are indicated with open green triangles. 
   Some representative evolutionary tracks of non-rotating models, for different initial masses (indicated along the tracks in units of solar mass), are also shown with solid coloured lines.
   }
\label{fig:densinv_grid}
\end{center}
\end{figure}

\begin{figure}
\begin{center}
\resizebox{\hsize}{!}{\mbox{\includegraphics[width=\linewidth, angle=-90]{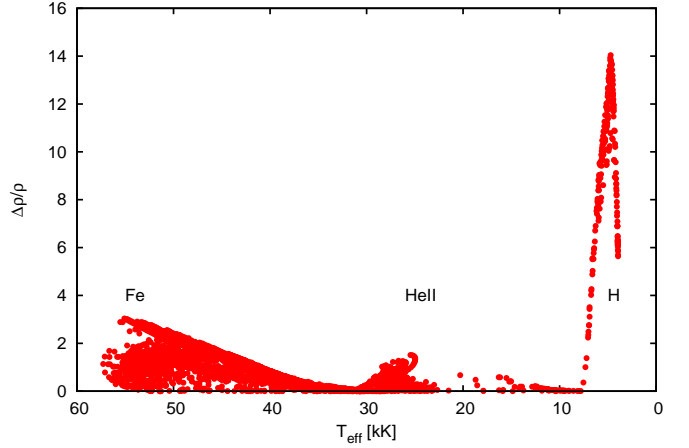}}}
   \caption[]{
     The extent of density inversions ($\Delta\rho/\rho$) as a function of $\mathrm{T_{eff}}$ for our models. The three 
     peaks correspond to the three opacity bumps of Fe, HeII and H in the OPAL tables, as indicated.} 
     \label{fig:densinv_Teff}
\end{center}
\end{figure}

Figure \ref{fig:densinv_grid} identifies our core hydrogen burning 
models which contain a density inversion. 
The quantity $\Delta\rho/\rho$ represents the strength 
of the density inversion normalized 
to the minimum density attained in the inflated zone.
We can identify three peaks in $\Delta\rho/\rho$ at 
$T_{\mathrm{eff}}/\mathrm{kK} \sim$ 55, 25 and 5.5 (see also Fig.~\ref{fig:densinv_Teff}),
which coincides exactly with the three $T_{\mathrm{eff}}$-regimes in which models
exceed the Eddington limit (cf., Fig.\,\ref{fig:gamma_Teff}).
The maximum of the density inversions in the three zones is related to the relative prominence 
of the three opacity bumps of Fe, HeII and H respectively, as shown in Fig.~\ref{fig:densinv_Teff}. 

\begin{figure}
 \begin{center}	
 \resizebox{\hsize}{!}{\mbox{\includegraphics[width=\linewidth, angle=-90]{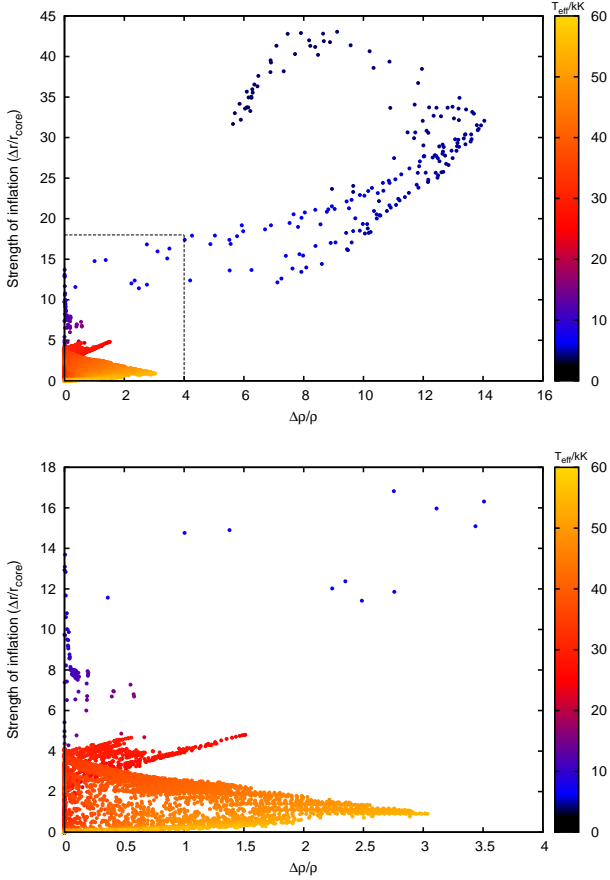}}}
    \caption[]{\textit{Top:} Correlation between inflation and density inversions for our models with $\log L/L_{\sun}>5$. 
    \textit{Bottom:} Zoomed-in view  of the area within the back dotted lines in the top panel. }
 \label{fig:densinv_inflation}
 \end{center}
 \end{figure}

However, an inflated model is not necessarily accompanied by
a density inversion. This is depicted clearly in Fig.~\ref{fig:densinv_inflation} 
where we investigate the correlation between inflation and density inversion 
(this can also be seen by comparing  
Fig.~\ref{fig:inflation_grid} to Fig.~\ref{fig:densinv_grid}). 
Figure \ref{fig:densinv_inflation} shows many models
which are even substantially inflated but do not develop a density inversion.  
The three peaks in the distribution of density inversions of Fig.~\ref{fig:densinv_Teff} 
also show up distinctly in this plot 
at the three characteristic effective temperatures (shown in color).
Models which do show a density inversion do always show some inflation. This is less obvious
from Fig.~\ref{fig:densinv_inflation}, because the hottest models show the smallest amount of
inflation (Fig.~\ref{fig:inflation_Teff}). 

The stability of density inversions in stellar envelopes 
has been a matter of debate for the 
last few decades but there has been no consensus on this issue 
yet \citep[see][]{maeder92}.  
There have been early speculations by \citet{mihalas69} 
while studying red supergiants that a density inversion might lead to Rayleigh-Taylor 
instabilities (RTI) resulting in ``elephant trunk'' structures washing out the 
positive density gradient. 
However, as rightly pointed out by \citet{schaerer96}, RTI will not 
develop since the effective gravity $g_{\mathrm{eff}}\, =g(1-\Gamma)$ acting on the 
fluid elements is directed outwards in the super-Eddington layers which contain the 
density inversion. \citet{kutter70} on the other hand claimed that a hydrodynamic treatment of 
the stellar structure equations will prevent any density inversion and would 
instead lead to a steady mass outflow. 
However, this claim is refuted by the present work, since our code 
solves the 1-D hydrodynamic stellar structure equations, in agreement 
with previous hydrodynamical 
models by \citet{glatzel&kiriakidis_93b} and \citet{meynet92}. 
\citet{stot_chin1973} suggested that density inversions will lead to strong turbulent motions 
instead of drastic mass loss episodes. However, these layers are unstable to convection, so 
turbulence is present in any case.

Additionally, \citet{glatzel&kiriakidis_93b} argued in favor of a sustainability of 
density inversions in the sense that they can be viewed as a natural consequence 
of strongly non-adiabatic convection, and they pointed out that the only 
plausible way to suppress density inversions is to use a different theory of convection. 
The only instability expected from simple arguments therefore is 
convection which is in line with \citet{wentzel1970} and \citet{langer97}.

Still, \citet{ekstroem2012} and \citet{yusof2013} recently considered density inversions 
as `unphysical'. Density inversions have been suppressed in their models 
by replacing the pressure scale height in the Mixing-Length Theory 
with the density scale height 
(cf. Sect.\,\ref{sec:comparison_lit}), as done by many stellar modelers in the past,  
often to prevent numerical difficulties. As the density scale height 
tends to infinity when a density inversion starts to develop, this measure
tends to enormously increase the convective flux in the relevant layers.
It is doubtful whether in reality the convective flux can be increased so much, as the
ratio of the local thermal to the local dynamical time scale in the relevant layers
is much smaller than one, such that convective eddies lose their thermal energy much 
faster than they rise, and thus hardly transport any energy at all.
Multi-dimensional hydrodynamical simulations are desirable to settle this issue.
We briefly return to this point in Sect.~\ref{sec:comparison_lit}.
 
\section{Sub-surface convection}\label{sec:convective_velocity}

We also studied the convective velocities 
in the sub-surface convection zones associated with the opacity peaks
in our stellar models \citep{cantiello2009}.  
We measure these velocities in units of either the
isothermal or the adiabatic sound velocity, i.e.
$c_{\deb{s,ad}}$ and $c_{\deb{s,iso}}$ respectively, which we compute as
\begin{equation}
 c_{\rm s,ad}=\sqrt{\frac{\gamma P}{\rho}} 
\end{equation} and
\begin{equation}
   c_{\rm s,iso}=\sqrt{\frac{k_{B}T}{\mu}} = \sqrt{\frac{P_{\deb{gas}}}{\rho}}, 
\end{equation}
where $\gamma$ is the adiabatic index,
$P$ is the total pressure, $\rho$ is the density,
$\mu$ is the mean molecular weight, $T$ is the temperature
and $k_{B}$ is the Boltzmann constant.
We define $M_{\rm iso}$ as the maximum ratio of the convective velocity
over the isothermal sound speed in the stellar envelope,
and $M_{\deb{ad}}$ correspondingly using the adiabatic sound speed.

\begin{figure}
\begin{center}	
 \resizebox{\hsize}{!}{\mbox{\includegraphics[width=\linewidth, angle=-90]{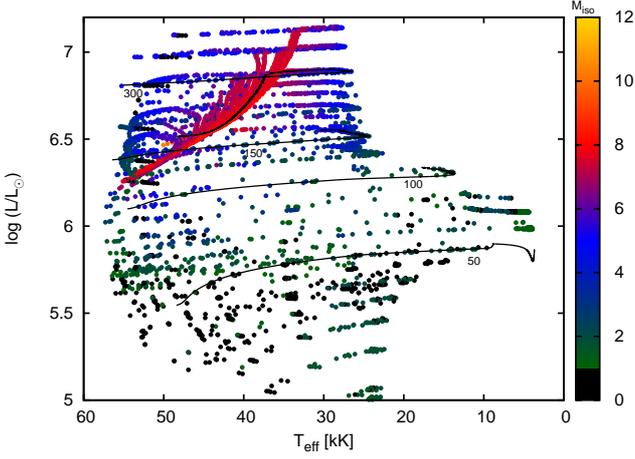}}}
   \caption[]{Upper HR diagram ($\log L/L_{\sun}>5$) 
   showing the maximum of the ratio of the convective velocity to the 
   local isothermal sound speed ($M_{\deb{iso}}$) of the analysed 
   models as coloured dots. Models with $M_{\deb{iso}}<1$ have been coloured black. 
   Some representative evolutionary tracks of non-rotating models, for different initial masses (indicated along the tracks in units of solar mass), are also shown with solid black lines.
   }
\label{fig:vconv_MS}
\end{center}
\end{figure}

\begin{figure}
\begin{center}	
 \resizebox{\hsize}{!}{\mbox{\includegraphics[width=\linewidth, angle=-90]{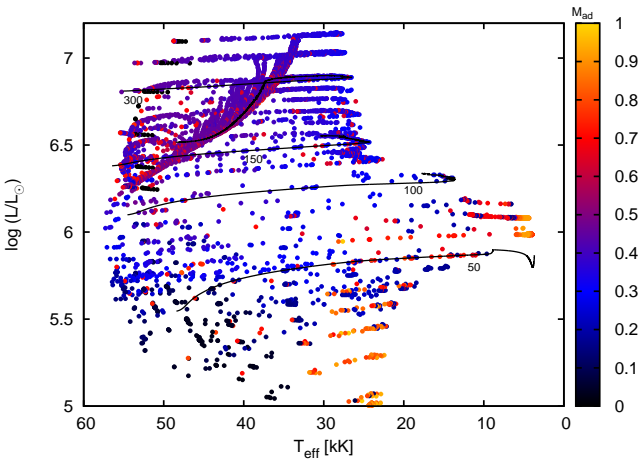}}}
   \caption[]{Upper HR diagram ($\log L/L_{\sun}>5$) 
   showing the maximum of the ratio of the convective velocity to the 
   local adiabatic sound speed $M_{\deb{ad}}$ of the analysed models 
   (computed with an adiabatic index of $5/3$) as coloured dots. 
   Some representative evolutionary tracks of non-rotating models, for different initial masses (indicated along the tracks in units of solar mass), are also shown with solid black lines.
   }
\label{fig:vconv_MS_ad}
\end{center}
\end{figure}

The true sound speed will be in between the adiabatic and isothermal one,
closer to the first one in the inner parts of the star, and closer to
the second in the inflated stellar envelope (cf., Sect.\,5).
In Figs.\,\ref{fig:vconv_MS} and~\ref{fig:vconv_MS_ad}, we show 
the values of $\deb{M_{iso}}$ and $M_{\deb{ad}}$ for our models
in the HR\,diagram. Whereas the convective velocities are always
smaller than the adiabatic sound speed, Fig.\,\ref{fig:vconv_MS} shows
that the isothermal sound speed can be exceeded locally in our models by a factor
of a few. The convective velocity and sound speed profiles for an extreme 
model are presented in Appendix \ref{app:supersonic}.

Supersonic convective velocities (adiabatic or isothermal, depending on 
the physical conditions in the envelope) may not be realistic and are outside the frame of 
the standard Mixing Length Theory. Therefore, in some of our models, the
convective velocities, and thus the convective energy transport, may have been
overestimated. A limitation of the velocities to the adequate sound speed
is expected to reduce the convective flux, which might lead to further inflation of the stellar envelope.

The cool models which have the strongest inflation have relatively  
smaller values of $M_{\deb{iso}}$ (compared to the hot WR-type models) 
but large values of $M_{\deb{ad}}$ (Fig.~\ref{fig:vconv_MS_ad}) in the 
sub-surface convection zones. This is primarily because of the fact 
that while $c_{\deb{s,ad}}$ depends on the total pressure, $c_{\deb{s,iso}}$ depends on the gas pressure only. 
In the very outer layers of the cool, luminous models, $\beta\rightarrow 1$ 
and hence $P_{\deb{gas}}\approx P_{\deb{tot}}$. In such situations, 
$c_{\deb{s,ad}}$ and $c_{\deb{s,iso}}$ are only a factor $\sqrt{\gamma}$ apart, where 
$\gamma$ is the adiabatic index.

\begin{figure}
  \begin{center}
  \resizebox{\hsize}{!}{\mbox{\includegraphics[width=\linewidth, angle=-90]{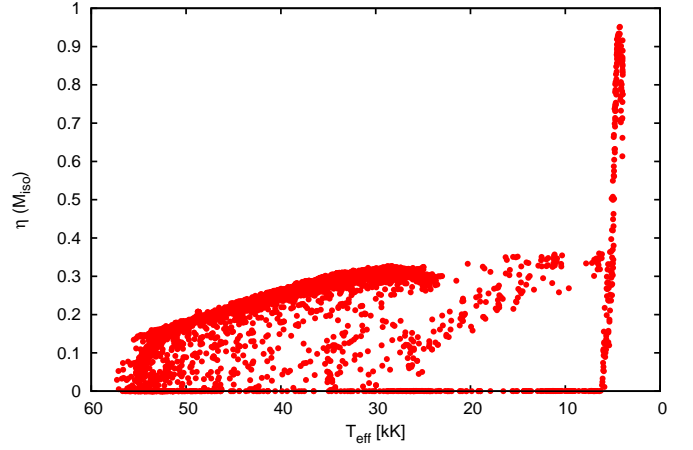}}}
  \caption[]{Convective efficiency $\eta(M_{\rm iso})$, which is the ratio of the convective flux to the total flux at the 
   position where the isothermal Mach number is the largest in the stellar envelope, as a function of the 
   effective temperature for all analysed stellar models in our grid.}
 \label{fig:conv_eff_temp}
 \end{center}
 \end{figure}

\begin{figure}
\begin{center}	
 \resizebox{\hsize}{!}{\mbox{\includegraphics[width=\linewidth, angle=-90]{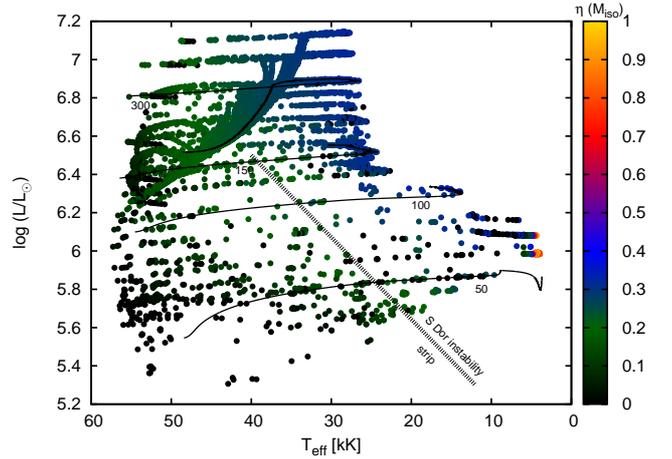}}}
   \caption[]{Upper HR diagram showing all the inflated stellar models in our grid as coloured dots. The convective efficiency 
   $\eta(M_{\rm iso})$, which is the ratio of the convective flux to the total flux at the 
   position where the isothermal Mach number is the largest in the stellar envelope, has been color coded. 
   Some representative evolutionary tracks of non-rotating models, for different initial masses (indicated along the tracks in units of solar mass), are also shown with solid black lines.
   }
\label{fig:conv_eff_HR}
\end{center}
\end{figure}

We find that the convective energy transport is not always negligible 
in the inflated models (cf. Sec \ref{subsec:discussion}).
We therefore evaluate the amount of flux that is actually carried by convection 
in the inflated envelopes of our models. 
We define the quantity $\eta(M_{\rm iso})$ as the fraction of the total flux carried by convection  
in the stellar envelope, at the location where 
the isothermal Mach number is the largest.
This quantity is plotted as a function of the 
effective temperature in Fig.~\ref{fig:conv_eff_temp}. 
It is evident from this figure that  $\eta(M_{\rm iso})$ needs not to be small 
for stellar envelopes to be inflated. However, the hotter a model is the lower its
 $\eta(M_{\rm iso})$-value at a given luminosity (see Fig.~\ref{fig:conv_eff_HR}).  
 For models hotter than 
$T_{\mathrm{eff}}\approx 63$ kK (for e.g. the hydrogen-free He stars), 
$\eta(M_{\rm iso})$ indeed goes towards zero (Grassitelli et al., in preparation). 
The behaviour of the quantity  $\eta(M_{\rm iso})$ in the HR\,diagram is 
shown in Fig.~\ref{fig:conv_eff_HR}.

\section{Comparison with previous studies}\label{sec:comparison_lit}

\subsection{Stellar atmosphere and wind models}

Since the Eddington limit was thought to be reached in massive stars near their surface
(cf., Sect.\,\ref{sec:edd_limit}), several papers have investigated this using stellar atmosphere calculations.
\citet{lamers_fitzpatrick_1988} investigated the Eddington factors in the atmospheres
of luminous stars in the temperature-gravity diagram, while \citet{ulmer_fitz98}
did so in the HR diagram. Both studies took the full radiative opacity into account.  
While their technique did not allow them to reach Eddington factors of one or more, Ulmer and Fitzpatrick
found that model atmospheres with a maximum Eddington factor of $0.9$ are located near the
observed upper luminosity limit of stars in the Large Magellanic Cloud \citep{hd1979,fitz_gar_1990}.

The main feature in the lines of constant Eddington-factors in the HR diagram found by Ulmer and Fitzpatrick 
is a drop from $T_{\deb{eff}} \simeq 60\,000\,$K to $15\,000\,$K. This may correspond to the drop in 
the maximum Eddington factor seen in our models in the same temperature interval (cf., Figs.\,\ref{fig:gamma_HR}
and \ref{fig:gamma_sHRD}).
Note that the peak around $T_{\deb{eff}} \simeq 30\,000\,$K in Fig.\,\ref{fig:gamma_Teff} corresponds only
to helium-rich models, which are not considered by Ulmer and Fitzpatrick.

On the other hand, neither inflation nor super-Eddington 
layers or density inversions are reported 
by Ulmer and Fitzpatrick, or from any hot, main sequence star
model atmosphere calculation so far (to the best of our knowledge). 
One reason might be that many model atmospheres 
only include a rather limited optical depth range
(e.g., up to $\tau=100$ in Ulmer and Fitzpatrick), such that 
the iron opacity peak is often not included
in the model. Additionally, the computational methods 
employed might not allow for a non-monotonic density profile.

Given the ubiquity of inflation for models above 
$\llso > 5.5$ or $M > 50\mso$ in the LMC, and a
correspondingly lower limit in the Milky Way 
due to its higher iron content, 
it is desirable to construct model atmospheres
which include this effect and identify its 
observational signatures. As the density profiles of 
such atmospheres near the photosphere are 
significantly different from those in non-inflated atmospheres,
such signatures may indeed be expected. 

\citet{asplund_1998} gives a thorough analysis of the Eddington limit in cool star atmosphere models. 
He does indeed find super-Eddington layers and density inversions in his models, and gives arguments for
the physically appropriate nature of these phenomena. He also discusses the effects of stellar winds on
these features, and finds they may be suppressed by extremely strong winds, but not by winds with mass loss rates
in the observed range. Asplund does not find inflation in his models, arguably because again, the iron opacity
peak is not included in his model atmospheres, which appears essential even for our models with cool effective 
temperatures.

\citet{owocki2004} and \citet{van_marle_2008} studied the winds of stars which
reach or exceed the Eddington limit at their surface. As we have shown above, this condition is generally not 
found in our models (cf., Figs.\,\ref{fig:surface_gamma_sHRD} and\,\ref{fig:gamma_surf_gammax}).
However, it may occur in helium-rich stars (see again Fig.\,\ref{fig:surface_gamma_sHRD}) and hydrogen-free
Wolf-Rayet stars \citep[cf.][]{heger_langer96}, as well as in stars which deviate from 
thermal or hydrostatic equilibrium. Noticeably, \citet{owocki2004} find that the mass loss rates in this case
are still quite limited, due to the energy loss attributed to lifting the wind material out of the
gravitational potential \citep[see,][]{heger_langer96}. 

We want to emphasize in this context that the Eddington limit 
investigated in the quoted
models as well as in our own may be different from 
the true Eddington limit, due to a number of effects
which are all related to the opacity of the 
stellar matter in the stellar envelopes.
One is that convection, which is necessarily 
present in the layers near or above the Eddington limit, 
may induce density inhomogeneities or clumping
which can alter the effective radiative opacity
\citep{shaviv1998}. In fact, depending on the 
nature of the clumping, the opacity may
be enhanced \citep{graefener_2012} or reduced 
\citep{owocki2004,rusz_2003,muijres_2011}. 
Furthermore, such opacity calculations are 
tedious, and even in the currently
used opacities, important contributions 
might still be missing. 

Finally, the effect of stellar rotation on the stability limit 
in the atmospheres especially of hot stars is clearly
important \citep{langer97,langer98,mm_2000}. 
However, it adds another dimension to this difficult problem
and is therefore generally not included (cf. Sect. \ref{subsec:interior_models}).

\subsection{Stellar interior models}\label{subsec:interior_models}

The peculiar core-halo density structure of 
inflated stars has first been pointed out
by \citet{stot_chin1993}, after the 
large iron bump in the opacities
near 170\,000\,K was found by \citet{ir92}. 
Further studies pointing out this phenomenon comprise 
\citet{ishii99}, \citet{petrovic_2006}, 
\citet{graefener_2012} and \citet{koehler2015}.
Conceivably, inflation may be present in 
further models of very massive stars, but often no statements
on the presence or absence of this phenomenon 
are made in the respective papers. 

For example, the models for very massive stars by \citet{yusof2013} 
only discuss the electron-scattering Eddington factor 
in their models. On the ZAMS, their models are hotter and more compact 
than the ones of \citet{koehler2015} 
analysed here, which implies that inflation is either weaker or absent. 
This difference might be due to the different 
treatment of convection in the sub-surface 
convective zones, where \citet{yusof2013} assume the
mixing length to be proportional to the 
density scale height instead of the standard pressure scale height.   
This prohibits the formation of density inversions, 
and since the density scale height tends to infinity when a model
attempts to establish a density inversion, 
the convection may transport an arbitrarily large 
energy flux in this scheme. While the physics of 
convection introduces one of the biggest
uncertainties in the atmospheres of stars close 
to the Eddington limit, efficient convective energy transport
in inflated envelopes appears unlikely (cf. Sect. \ref{sec:dens_inv}).

A suppression of inflation may have significant 
consequences for the evolution of massive stars,
as the stellar models stay bluer and as a 
result have lower mass loss rates and 
lower spin-down rates. The final fates 
of such non-inflated stars will be significantly 
different compared to inflated stars 
\citep[see][for a detailed discussion]{koehler2015}.

\citet{graefener_2012} find inflation which, for their models without clumping,
correspond well to those of \citet{petrovic_2006} for the Wolf-Rayet case,
and to our unpublished solar metallicity main sequence models,
which show a bending of the zero-age main sequence to cool temperatures for
$M\simgr 100\mso$. The models of \citet{ishii99} again agree very well. 
Including the work of \citet{stot_chin1993}, we
conclude that the effect of inflation in models of massive main sequence stars
is found in at least four independent stellar structure codes, with
three of them quantitatively producing very similar results.

As pointed out above, massive star evolutionary 
models which include effects of rotation
are being produced routinely these days 
\citep[cf.,][]{maeder_2010,langer2012,chieffi_2013},
but an investigation of the effect of stellar rotation 
on the stability limit in the atmospheres of hot stars
requires the construction of two-dimensional stellar models.


\section{Comparison with observations}\label{sec:comparison_obs}

\subsection{The VFTS sample}

\begin{figure*}
\begin{center}
\resizebox{\textwidth}{!}{\includegraphics[angle=-90]{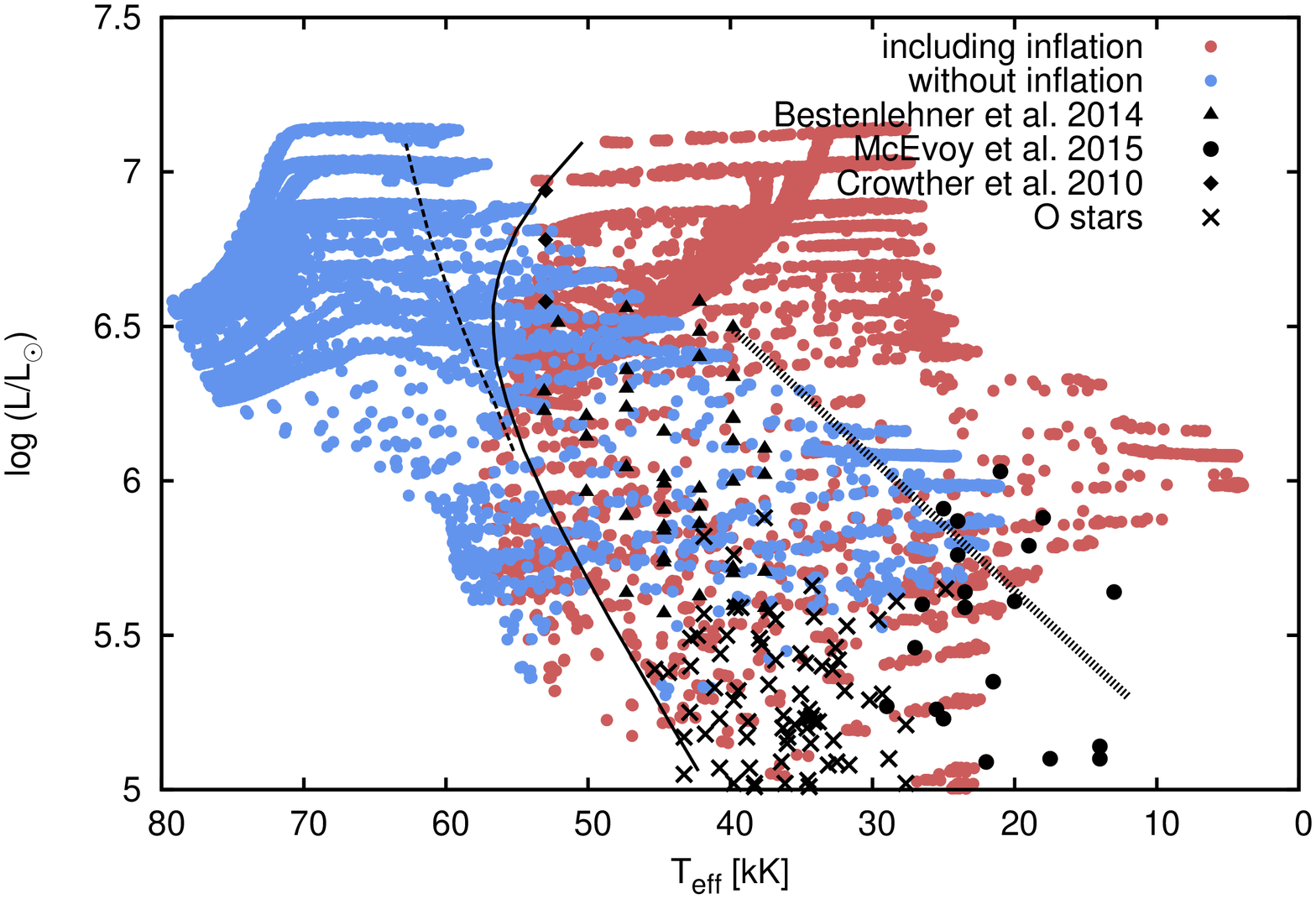}}
    \caption[]{Hertzsprung-Russell diagram showing the K\"oehler et al. models which include 
    inflation (red) and corresponding 
    non-inflated stellar models (blue) above $\log L/L_{\sun}>5$ (see text for explanation). 
    The triangles refer to all the Of/WN and WNh single stars  
    studied by \citet{bestenlehner2014} whereas the black dots 
    and the diamonds refer to the B-supergiants (single) from \citet{mcevoy_2015} and 
    WN5h stars of the core R136 from \citet{crowther2010}, respectively. 
    The O\,stars observed within the VFTS survey are 
     marked with crosses \citep[][Ramirez-Agudelo et al. in preparation]{carolina_2014}. 
     The zero-age main-sequence of 
     the non-rotating stars is marked with the solid line while 
     the dashed line indicates the approximate position of the zero-age main 
     sequence if inflation was absent (see text for further details). 
     The hot part of the S\,Doradus instability strip 
     from \citet{smith2004} is also shown for reference. 
     }
\label{fig:inflation_compare_Joachim}
\end{center}
\end{figure*}

A prime motivation of \citet{koehler2015} for computing the evolutionary
models for the very massive stars analysed here was to provide a theoretical framework
for the VLT~Flames-Tarantula Survey  \citep[VFTS,][]{evans2011}.
Within VFTS, multi-epoch spectral data of about 700 early B and 300 O~stars are
being analysed through detailed model atmosphere calculations. Within this
effort, \citet{bestenlehner2014} and \citet{mcevoy_2015} derived the physical
properties of more than 50 very massive stars, with luminosities 
$\log (L/\lso) > 5.5$. We confront the models of \citet{koehler2015} with this sample in
Fig.~\ref{fig:inflation_compare_Joachim}. 

Two sets of model data are included in Fig.~\ref{fig:inflation_compare_Joachim},
one which uses the effective temperatures of the K\"ohler et al. models directly,
and a second one where the effective core temperature 
is used as defined in Sect.\,\ref{sec:inflation}
(cf. Eq.\,\ref{eq:teff_core}). The latter approximates 
the surface temperature of our models if
inflation was completely absent. An example calculation presented in Appendix \ref{app:mix_length_comparison},
where inflation in a $300\mso$ is suppressed by increasing the mixing length parameter, 
shows that this approximation is indeed quite good. The zero-age main sequence 
is also drawn for both sets of models. Note that while 
wind effects are clearly seen in the spectra of all
stars in the sample, the optical depth of their winds 
is expected not to exceed $\tau=2$ (cf., fig.\,7 in 
\citet{koehler2015}) until the stars become very helium rich
at their surface. Therefore, the effective temperatures 
derived from the observations 
need not be corrected for optically thick winds.

As shown in \citet{bestenlehner2014}, the hottest stars in their sample follow
the ZAMS of the K\"ohler et al. models very closely, well into the regime of
inflation. 
One might expect un-evolved stars to the left of the K\"ohler et al. ZAMS if inflation
was not present. In that case, the stars above $\log L/L_{\sun} \simeq 6.2$ might spend 
a significant fraction of their life time on the hot side of the K\"ohler et al. ZAMS.
The absence of such hot stars, however, does not conclusively argue that inflation does exist in
nature, i.e., even without inflation, the star formation history in 30\,Doradus might preclude the 
existence of such stars, or they may be hidden in their natal cloud due to their youth
\citep{yorke_1986,castro_2014}.

On the cool side, Fig.~\ref{fig:inflation_compare_Joachim} shows an absence of observed stars
for $\log L/L_{\sun} \gtrsim 6.15$ and $T_{\rm eff} \simle 35\,000\,$K. As the evolutionary models
predict about 30\% of the core hydrogen burning to take place at $T_{\rm eff} \simle 35\,000\,$K
in this luminosity regime, this may indicate that the inflation in the models of K\"ohler et al.
is too strong. On the other hand, again, the absence of correspondingly cool stars 30\,Doradus
may be a result of the local star formation history. 

In the luminosity range below, at $5.5 \lesssim\log L/L_{\sun} \lesssim 6.15$, stars 
as cool as $T_{\rm eff} \simle 15\,000\,$K are observed for which \citet{mcevoy_2015} concluded
that they are still core hydrogen burning objects. The observed stars are somewhat cooler than
the coolest core effective temperature of our models, which may argue in favor of inflation in
real stars. Note that the life time of stars in the regime $T_{\rm eff} < 20\,000\,$K is
only 10\% for the K\"ohler et al. models in the considered luminosity range. 
 
In summary, as the stellar evolution models for these high masses are still quite uncertain,
we find it not possible to argue for or against inflation being present in the observed stars considered
here based on Fig.~\ref{fig:inflation_compare_Joachim}. In fact, it is intriguing that most
of the observed stars are found in the regime where the inflated and non-inflated models overlap.
Nevertheless, the observed sample above $\log L/\lso\simeq 5.5$ might constitute the best test case, 
since according to our models, the envelopes of all of them 
are expected to be strongly affected by the Eddington limit.
Model atmosphere calculations for these stars which include 
inflation might shed new light on this question.

\subsection{Further possible consequences of inflation}\label{subsec:LBV}

\subsubsection*{S\,Doradus type variability} 
\citet{graefener_2012} argue that the S\,Doradus type variability of 
LBVs may be related to the effect of inflation, and focussed in particular
on the case of AG Car \citep{groh2009}. They propose
that an instability sets in when their $70\, \mathrm{M_{\sun}}$
chemically homogeneous hydrostatic stellar model is
highly inflated ($\approx 120\,\mathrm{R_{\sun}}$) by virtue
of which the inflated layer becomes gravitationally unbound and a
mass loss episode follows.

In contrast to this idea, our inflated hydrodynamic stellar models do not show any
signs of such an instability. This could be due to various simplifying assumptions
made in the models of \citet{graefener_2012}, in particular their neglect of
the convective flux in the inflated envelopes 
(cf., Sect.\ref{sec:convective_velocity}). Nevertheless, the physics
of convection is very complex in these envelopes, and our
results do not imply that instabilities may not occur. 

In fact, considering the hot edge of the S\,Doradus variability strip according to \citet{smith2004}
in Fig.~\ref{fig:conv_eff_HR}, we see that it roughly separates the models with
a low maximum convective efficiency ($\eta_{\rm max} < 0.2$) from those
with a higher convective efficiency. If such high fluxes would not be achievable in these
envelopes (cf., Sect. \ref{sec:convective_velocity}), a dynamical instability might well be possible.

The hot edge of the S\,Doradus variability in the 
HR diagram also coincides quite well with the borderline
separating mildly ($\Delta r/r_{\rm core} < 1$) 
from strongly ($\Delta r/r_{\rm core} > 1$)
inflated models. Comparing this with the observed 
distribution of very massive stars in
Fig.~\ref{fig:inflation_compare_Joachim}, 
which indicates that essentially no stars
are found far to the cool side of this line, could 
indicate again that strongly inflated
envelopes are indeed unstable, and might 
lead to S\,Doradus type variability and an increased 
time-averaged mass loss rate.

\subsubsection*{LBV eruptions} 

\citet{glatzel&kiriakidis_93b} speculated that strange mode pulsations
might be responsible for the LBV phenomenon. These pulsations are
characterized by very short growth times ($\sim \tau_{\mathrm{dyn}}$)
and small brightness fluctuations roughly of the order
$\sim 10\ldots 100$ mmag \citep{glatzel99,grott2005}.
However, the mass contained in the pulsating
envelopes of their models is negligible compared to the stellar mass, and the associated brightness
variations cannot explain the humongous
luminosity variations observed in LBV eruptions.

We have seen in Sect. \ref{sec:dens_inv} that an 
inflated envelope often produces a density inversion.
Such density inversions have been repeatedly 
proposed as a source of instabilities giving rise to eruptive
mass loss in LBVs \citep{maeder89,maeder94,stot_chin1993}. 
Given our results, we consider it unlikely that a density inversion
can be the sole cause of LBV eruptions. 
Density inversions are a generic feature present 
in a multitude of our models (see Fig.~\ref{fig:densinv_grid}),
while the LBV phenomenon is quite rare. 
Furthermore, the density inversions in our models
are found very close to the surface of the star,
with very small amounts of mass above it.

\begin{figure}[H]
\begin{center}	
\resizebox{\hsize}{!}{\includegraphics[angle=-90]{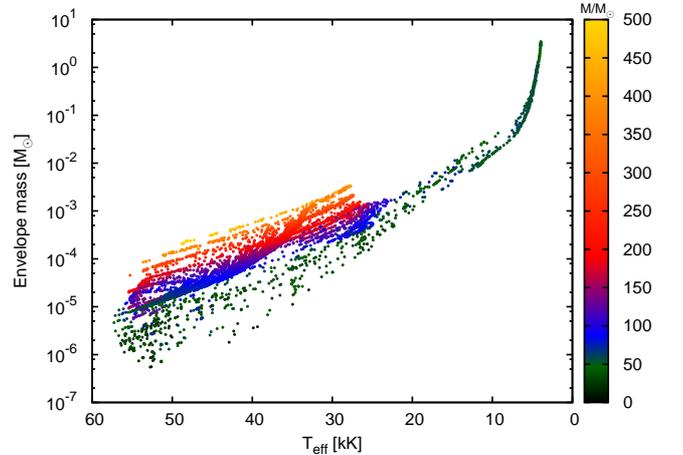}}
   \caption[]{   
   Mass contained in the inflated envelope for all inflated models of our grid, 
   as a function of the effective temperature. The effective actual mass of the models is 
   colour-coded (see colour bar to the right). 
   }
\label{fig:envmass}
\end{center}
\end{figure}

Given our results, inflation per se appears unlikely to cause LBV eruptions, again,
because it occurs too abundantly in our models, and also because the mass of the inflated
envelope is generally very small. However,  Fig.\,\ref{fig:envmass} reveals that this is
not so for our coolest models. Whereas for most models the mass of the inflated envelope 
is smaller than $\sim 10^{-3}\mso$, intriguingly it rises to 
several solar masses in the models which have effective temperatures below
$\sim 10\,000\,$K. These cool models, of which detailed examples are
presented in Fig.\,\ref{fig:app:H-bump}, also show the highest Eddington factors (Fig.\,\ref{fig:gamma_Teff}) and the 
strongest inflation (Fig.\,\ref{fig:inflation_Teff}). This behaviour is seen in the mass range of $40\dots 100\mso$, 
which corresponds well to the masses of observed LBVs.

A key feature in our cool models with massive inflated envelopes is visible in Fig.\,\ref{fig:conv_eff_temp}.
As the opacity in the hydrogen recombination zone becomes very large, effectively blocking any
radiation transport, convective efficiencies of the order of $\eta\simeq 1$ are needed 
to transport the stellar luminosity through this zone. Also, in the iron convection zone at
the bottom of the envelope, a high convective efficiency ($\eta\simeq 0.5$) is found in these
models.  As shown in Figs.\,\ref{fig:vconv_MS} and \ref{fig:vconv_MS_ad},
this requires sonic or even supersonic convective velocities in the framework of the standard
MLT as implemented in our code. It thus appears conceivable that in reality convection is less
efficient in such a situation, for e.g., because of viscous dissipation.
When a star enters this region with a massive inflated envelope,
throughout which the stellar luminosity can neither be transported by radiation
nor by convection, hydrostatic equilibrium will not be possible any more, and the 
loosely bound inflated envelope may by dynamically ejected. We believe that this 
scenario may relate to major LBV eruptions.

As proposed by \citet{langer2012}, a rapid evolutionary time scale of a star
may be required to obtain LBV outbursts in addition to the star reaching
the Eddington limit. If evolution on the thermal timescale is rapid enough,
this might produce LBVs after core hydrogen exhaustion which may relate to most
of the observed LBVs in our Galaxy and the Magellanic Clouds, as well as LBVs after core helium exhaustion, which may
concern to the recently accumulated evidence of LBV outbursts in immediate
supernova progenitors \citep{smith+arnett2014}.

\subsubsection*{Supernova shock break-out} 
A recent study by \citet{moriya2015} concluded that inflated stellar 
models can help to explain the extended rise time of the shock break-out signal from the 
Type Ib supernova SN2008D \citep{soderberg2008}. In this scenario the shock break-out occurs deep inside the inflated envelope and 
consequently the rise time is determined by the radiative diffusion time of the envelope and not the light crossing time. 
They also noted that more such events, if observed in future, might serve as indicators of inflated 
supernova progenitors.

Whereas the above result supports the idea that hydrogen-free 
Wolf-Rayet stars may possess inflated envelopes
\citep{petrovic_2006,graefener_2012}, LBVs have also been 
suggested to be immediate progenitors of supernovae \citep{kotak_vink2006,groh2013}. 
In the realm of high cadence supernova surveys, 
this opens up the possibility to also test the existence of 
envelope inflation in hydrogen-rich stars through 
supernova shock break-out observations.


\section{Discussion and conclusions}\label{sec:conclusions}

We investigated the internal structures of the massive star models computed by \citet{ines2011}
and \citet{koehler2015} using a 1-D hydrodynamical stellar evolution code, with particular
emphasis to the Eddington limit.
We find that the conventional idea of sufficiently massive stars reaching the Eddington limit at the stellar
surface is not reproduced by our core hydrogen burning models, 
not even at 500$\mso$ (cf., Figs.\,\ref{fig:surface_gamma_sHRD} and \ref{fig:gamma_surf_gammax}). 
Instead, we find a suitably defined Eddington limit inside the star (Eqn.\,\ref{eq:gamma_new}) is 
reached by models with $\log (L/\lso) \gtrsim 5.6$ (Fig.\,\ref{fig:gamma_HR}),
which leads to sub-surface convection, envelope inflation (Fig.\,\ref{fig:inflation_gamma}) and possibly to pulsations.
  Many of our models even exceed this Eddington limit, in the extreme case of red supergiants
even by factors of up to seven (Fig.\,\ref{fig:gamma_Teff}), with the consequence that strong density inversions develop
such that hydrostatic equilibrium is maintained (Fig.\,\ref{fig:densinv_Teff}). 

In the analysed models, whose initial composition is chosen to match that of the LMC, all stars
above $\sim 40\mso$ do reach the Eddington limit in their envelopes. As iron opacities are mainly
responsible for this phenomenon, we expect that this mass limit is higher at a lower metallicity,
and similarly lower for massive stars in our galaxy. Furthermore, there
may be two groups of stars for which this limit comes down even further. Firstly, the centrifugal
force in rapidly rotating stars may lead to similar conditions in the envelope layers near the
stellar equator, i.e. to a strong latitude dependence of inflation. Perhaps, this could
give rise to the so called B[e] supergiants, which show a slow and dense equatorial wind and
a fast polar wind at the same time \citep{zickgraf_1985}. Second, the mass losing stars in interacting 
close binary systems evolve to much higher $L/M$-values than corresponding single stars \citep{langer2014}, and
are therefore expected to reach the Eddington limit for much lower initial masses.

The stability of the inflated envelopes is not investigated here, but many of them are likely to
be pulsationally unstable \citep[][Sanyal et al. in preparation]{glatzel&kiriakidis_93b,saio98}. 
If so, it is expected that the
pulsations will lead to mass loss enhancements \citep[e.g.][]{moriya_langer_2015}, or to the loss
of the inflated envelope. In the latter case, the envelope is expected to re-grow unless
the achieved time average mass loss rate exceeds the high critical mass loss rate 
(Sect.\,\ref{subsec:masloss_inflation}). 
We find that in our coolest models, the mass contained in the inflated envelopes can
reach several solar masses (Fig.\,\ref{fig:envmass}), and
speculate that their dynamical loss may resemble LBV major eruptions (cf.\, Sect.\,\ref{subsec:LBV}).
Consequently, even though reaching or exceeding the Eddington limit may not immediately lead to strong 
outflows in stars, clearly the mass loss rate of the stars will be strongly affected, in the 
sense that the mass loss will be significantly enhanced one way or another. 

It will be crucial to test observationally whether luminous, main sequence stars indeed possess 
inflated stellar envelopes. This possibility has not yet been investigated with stellar
atmosphere models for hot stars. Perhaps the best candidates are the S\,Doradus variables
\citep{graefener_2012}, which appear in the part of the HR diagram where our models
predict a radius inflation by more than a factor of two (cf., Fig.\,\ref{fig:inflation_grid}).

Finally, we note that besides massive stars, the Eddington limit is relevant to various other 
types of stars, as luminous post-AGB star, X-ray bursts, Novae, R\,Corona Borealis stars and 
accreting compact objects. It may be interesting to assess to what extent similar phenomena as
found in this work might play a role in these objects.  

\begin{acknowledgements}
We thank the referee, Raphael Hirschi, for useful comments that improved 
the manuscript. We are grateful to S.\,C.\,Yoon for many fruitful discussions. We also thank 
C.\,J.\,Evans, D.\,Sz\'ecsi and J.\,S.\,Vink 
for helpful comments on the manuscript. L.\,G. thanks the 
International Max Planck Research School for Astronomy and Astrophysics at
Bonn.
  \end{acknowledgements}

\bibliographystyle{aa}
\bibliography{debashis_refs}

\begin{thebibliography}{100}
\expandafter\ifx\csname natexlab\endcsname\relax\def\natexlab#1{#1}\fi

\bibitem[{{Asplund}(1998)}]{asplund_1998}
{Asplund}, M. 1998, \aap, 330, 641

\bibitem[{{Bestenlehner} {et~al.}(2014){Bestenlehner}, {Gr{\"a}fener}, {Vink},
  {Najarro}, {de Koter}, {Sana}, {Evans}, {Crowther}, {H{\'e}nault-Brunet},
  {Herrero}, {Langer}, {Schneider}, {Sim{\'o}n-D{\'{\i}}az}, {Taylor}, \&
  {Walborn}}]{bestenlehner2014}
{Bestenlehner}, J.~M., {Gr{\"a}fener}, G., {Vink}, J.~S., {et~al.} 2014, \aap,
  570, A38

\bibitem[{{B{\"o}hm-Vitense}(1958)}]{MLT1958}
{B{\"o}hm-Vitense}, E. 1958, \zap, 46, 108

\bibitem[{{Braun}(1997)}]{bra97}
{Braun}, H. 1997, PhD thesis, , Ludwig-Maximilians-Univ.~M{\"u}nchen, (1997)

\bibitem[{{Bresolin} {et~al.}(2008){Bresolin}, {Crowther}, \&
  {Puls}}]{Bresolin_2008}
{Bresolin}, F., {Crowther}, P.~A., \& {Puls}, J., eds. 2008, IAU Symposium,
  Vol. 250, {Massive Stars as Cosmic Engines}

\bibitem[{{Brott} {et~al.}(2011){Brott}, {de Mink}, {Cantiello}, {Langer}, {de
  Koter}, {Evans}, {Hunter}, {Trundle}, \& {Vink}}]{ines2011}
{Brott}, I., {de Mink}, S.~E., {Cantiello}, M., {et~al.} 2011, \aap, 530, A115

\bibitem[{{Cantiello} {et~al.}(2009){Cantiello}, {Langer}, {Brott}, {de Koter},
  {Shore}, {Vink}, {Voegler}, {Lennon}, \& {Yoon}}]{cantiello2009}
{Cantiello}, M., {Langer}, N., {Brott}, I., {et~al.} 2009, \aap, 499, 279

\bibitem[{{Castro} {et~al.}(2014){Castro}, {Fossati}, {Langer},
  {Sim{\'o}n-D{\'{\i}}az}, {Schneider}, \& {Izzard}}]{castro_2014}
{Castro}, N., {Fossati}, L., {Langer}, N., {et~al.} 2014, \aap, 570, L13

\bibitem[{{Chieffi} \& {Limongi}(2013)}]{chieffi_2013}
{Chieffi}, A. \& {Limongi}, M. 2013, \apj, 764, 21

\bibitem[{{Crowther} {et~al.}(2010){Crowther}, {Schnurr}, {Hirschi}, {Yusof},
  {Parker}, {Goodwin}, \& {Kassim}}]{crowther2010}
{Crowther}, P.~A., {Schnurr}, O., {Hirschi}, R., {et~al.} 2010, \mnras, 408,
  731

\bibitem[{{de Jager}(1984)}]{deJager1984}
{de Jager}, C. 1984, \aap, 138, 246

\bibitem[{{Doran} {et~al.}(2013){Doran}, {Crowther}, {de Koter}, {Evans},
  {McEvoy}, {Walborn}, {Bastian}, {Bestenlehner}, {Gr{\"a}fener}, {Herrero},
  {K{\"o}hler}, {Ma{\'{\i}}z Apell{\'a}niz}, {Najarro}, {Puls}, {Sana},
  {Schneider}, {Taylor}, {van Loon}, \& {Vink}}]{doran_2013}
{Doran}, E.~I., {Crowther}, P.~A., {de Koter}, A., {et~al.} 2013, \aap, 558,
  A134

\bibitem[{{Eddington}(1926)}]{eddington_1926}
{Eddington}, A.~S. 1926, {The Internal Constitution of the Stars}

\bibitem[{{Ekstr{\"o}m} {et~al.}(2012){Ekstr{\"o}m}, {Georgy}, {Eggenberger},
  {Meynet}, {Mowlavi}, {Wyttenbach}, {Granada}, {Decressin}, {Hirschi},
  {Frischknecht}, {Charbonnel}, \& {Maeder}}]{ekstroem2012}
{Ekstr{\"o}m}, S., {Georgy}, C., {Eggenberger}, P., {et~al.} 2012, \aap, 537,
  A146

\bibitem[{{Endal} \& {Sofia}(1976)}]{endal_1976}
{Endal}, A.~S. \& {Sofia}, S. 1976, \apj, 210, 184

\bibitem[{{{\'E}rgma}(1971)}]{ergma1971}
{{\'E}rgma}, {\'E}. 1971, \sovast, 15, 51

\bibitem[{{Evans} {et~al.}(2011){Evans}, {Taylor}, {H{\'e}nault-Brunet},
  {Sana}, {de Koter}, {Sim{\'o}n-D{\'{\i}}az}, {Carraro}, {Bagnoli}, {Bastian},
  {Bestenlehner}, {Bonanos}, {Bressert}, {Brott}, {Campbell}, {Cantiello},
  {Clark}, {Costa}, {Crowther}, {de Mink}, {Doran}, {Dufton}, {Dunstall},
  {Friedrich}, {Garcia}, {Gieles}, {Gr{\"a}fener}, {Herrero}, {Howarth},
  {Izzard}, {Langer}, {Lennon}, {Ma{\'{\i}}z Apell{\'a}niz}, {Markova},
  {Najarro}, {Puls}, {Ramirez}, {Sab{\'{\i}}n-Sanjuli{\'a}n}, {Smartt},
  {Stroud}, {van Loon}, {Vink}, \& {Walborn}}]{evans2011}
{Evans}, C.~J., {Taylor}, W.~D., {H{\'e}nault-Brunet}, V., {et~al.} 2011, \aap,
  530, A108

\bibitem[{{Fitzpatrick} \& {Garmany}(1990)}]{fitz_gar_1990}
{Fitzpatrick}, E.~L. \& {Garmany}, C.~D. 1990, \apj, 363, 119

\bibitem[{{Fuller} {et~al.}(1986){Fuller}, {Woosley}, \&
  {Weaver}}]{fuller_1986}
{Fuller}, G.~M., {Woosley}, S.~E., \& {Weaver}, T.~A. 1986, \apj, 307, 675

\bibitem[{{Gal-Yam} {et~al.}(2009){Gal-Yam}, {Mazzali}, {Ofek}, {Nugent},
  {Kulkarni}, {Kasliwal}, {Quimby}, {Filippenko}, {Cenko}, {Chornock},
  {Waldman}, {Kasen}, {Sullivan}, {Beshore}, {Drake}, {Thomas}, {Bloom},
  {Poznanski}, {Miller}, {Foley}, {Silverman}, {Arcavi}, {Ellis}, \&
  {Deng}}]{gal-yam2009}
{Gal-Yam}, A., {Mazzali}, P., {Ofek}, E.~O., {et~al.} 2009, \nat, 462, 624

\bibitem[{{Glatzel} \& {Kiriakidis}(1993)}]{glatzel&kiriakidis_93b}
{Glatzel}, W. \& {Kiriakidis}, M. 1993, \mnras, 263, 375

\bibitem[{{Glatzel} {et~al.}(1999){Glatzel}, {Kiriakidis}, {Chernigovskij}, \&
  {Fricke}}]{glatzel99}
{Glatzel}, W., {Kiriakidis}, M., {Chernigovskij}, S., \& {Fricke}, K.~J. 1999,
  \mnras, 303, 116

\bibitem[{{Gr{\"a}fener} {et~al.}(2012){Gr{\"a}fener}, {Owocki}, \&
  {Vink}}]{graefener_2012}
{Gr{\"a}fener}, G., {Owocki}, S.~P., \& {Vink}, J.~S. 2012, \aap, 538, A40

\bibitem[{{Gr{\"a}fener} {et~al.}(2011){Gr{\"a}fener}, {Vink}, {de Koter}, \&
  {Langer}}]{graefener2011}
{Gr{\"a}fener}, G., {Vink}, J.~S., {de Koter}, A., \& {Langer}, N. 2011, \aap,
  535, A56

\bibitem[{{Groh} {et~al.}(2009){Groh}, {Hillier}, {Damineli}, {Whitelock},
  {Marang}, \& {Rossi}}]{groh2009}
{Groh}, J.~H., {Hillier}, D.~J., {Damineli}, A., {et~al.} 2009, \apj, 698, 1698

\bibitem[{{Groh} {et~al.}(2013){Groh}, {Meynet}, \& {Ekstr{\"o}m}}]{groh2013}
{Groh}, J.~H., {Meynet}, G., \& {Ekstr{\"o}m}, S. 2013, \aap, 550, L7

\bibitem[{{Grott} {et~al.}(2005){Grott}, {Chernigovski}, \&
  {Glatzel}}]{grott2005}
{Grott}, M., {Chernigovski}, S., \& {Glatzel}, W. 2005, \mnras, 360, 1532

\bibitem[{{Hamann} {et~al.}(1995){Hamann}, {Koesterke}, \&
  {Wessolowski}}]{hamann95}
{Hamann}, W.-R., {Koesterke}, L., \& {Wessolowski}, U. 1995, \aap, 299, 151

\bibitem[{{Heger} \& {Langer}(1996)}]{heger_langer96}
{Heger}, A. \& {Langer}, N. 1996, \aap, 315, 421

\bibitem[{{Heger} {et~al.}(2000){Heger}, {Langer}, \& {Woosley}}]{heger2000}
{Heger}, A., {Langer}, N., \& {Woosley}, S.~E. 2000, \apj, 528, 368

\bibitem[{{Humphreys} \& {Davidson}(1979)}]{hd1979}
{Humphreys}, R.~M. \& {Davidson}, K. 1979, \apj, 232, 409

\bibitem[{{Iglesias} \& {Rogers}(1996)}]{ir96}
{Iglesias}, C.~A. \& {Rogers}, F.~J. 1996, \apj, 464, 943

\bibitem[{{Iglesias} {et~al.}(1992){Iglesias}, {Rogers}, \& {Wilson}}]{ir92}
{Iglesias}, C.~A., {Rogers}, F.~J., \& {Wilson}, B.~G. 1992, \apj, 397, 717

\bibitem[{{Ishii} {et~al.}(1999){Ishii}, {Ueno}, \& {Kato}}]{ishii99}
{Ishii}, M., {Ueno}, M., \& {Kato}, M. 1999, \pasj, 51, 417

\bibitem[{{Joss} {et~al.}(1973){Joss}, {Salpeter}, \& {Ostriker}}]{joss73}
{Joss}, P.~C., {Salpeter}, E.~E., \& {Ostriker}, J.~P. 1973, \apj, 181, 429

\bibitem[{{Kato}(1986)}]{kato86}
{Kato}, M. 1986, \apss, 119, 57

\bibitem[{{Kippenhahn} \& {Thomas}(1970)}]{kipp_thomas_1970}
{Kippenhahn}, R. \& {Thomas}, H.-C. 1970, in IAU Colloq. 4: Stellar Rotation,
  ed. A.~{Slettebak}, 20

\bibitem[{{Kippenhahn} \& {Weigert}(1990)}]{kw90}
{Kippenhahn}, R. \& {Weigert}, A. 1990, {Stellar Structure and Evolution}
  (Springer, Berlin)

\bibitem[{{K{\"o}hler} {et~al.}(2015){K{\"o}hler}, {Langer}, {de Koter}, {de
  Mink}, {Crowther}, {Evans}, {Gr{\"a}fener}, {Sana}, {Sanyal}, {Schneider}, \&
  {Vink}}]{koehler2015}
{K{\"o}hler}, K., {Langer}, N., {de Koter}, A., {et~al.} 2015, \aap, 573, A71

\bibitem[{{Kotak} \& {Vink}(2006)}]{kotak_vink2006}
{Kotak}, R. \& {Vink}, J.~S. 2006, \aap, 460, L5

\bibitem[{{Kozyreva} {et~al.}(2014){Kozyreva}, {Yoon}, \&
  {Langer}}]{kozyreva_2014}
{Kozyreva}, A., {Yoon}, S.-C., \& {Langer}, N. 2014, \aap, 566, A146

\bibitem[{{Kudritzki} \& {Puls}(2000)}]{kudritzki2000}
{Kudritzki}, R.-P. \& {Puls}, J. 2000, \araa, 38, 613

\bibitem[{{Kutter}(1970)}]{kutter70}
{Kutter}, G.~S. 1970, \apj, 160, 369

\bibitem[{{Lamers} \& {Fitzpatrick}(1988)}]{lamers_fitzpatrick_1988}
{Lamers}, H.~J.~G.~L.~M. \& {Fitzpatrick}, E.~L. 1988, \apj, 324, 279

\bibitem[{{Langer}(1991)}]{lan91}
{Langer}, N. 1991, \aap, 252, 669

\bibitem[{{Langer}(1997)}]{langer97}
{Langer}, N. 1997, in Astronomical Society of the Pacific Conference Series,
  Vol. 120, Luminous Blue Variables: Massive Stars in Transition, ed. A.~{Nota}
  \& H.~{Lamers}, 83

\bibitem[{{Langer}(1998)}]{langer98}
{Langer}, N. 1998, \aap, 329, 551

\bibitem[{{Langer}(2012)}]{langer2012}
{Langer}, N. 2012, \araa, 50, 107

\bibitem[{{Langer} \& {Kudritzki}(2014)}]{langer2014}
{Langer}, N. \& {Kudritzki}, R.~P. 2014, \aap, 564, A52

\bibitem[{{Larsson} {et~al.}(2007){Larsson}, {Levan}, {Davies}, \&
  {Fruchter}}]{larsson2007}
{Larsson}, J., {Levan}, A.~J., {Davies}, M.~B., \& {Fruchter}, A.~S. 2007,
  \mnras, 376, 1285

\bibitem[{{Maeder}(1989)}]{maeder89}
{Maeder}, A. 1989, in Astrophysics and Space Science Library, Vol. 157, IAU
  Colloq. 113: Physics of Luminous Blue Variables, ed. K.~{Davidson}, A.~F.~J.
  {Moffat}, \& H.~J.~G.~L.~M. {Lamers}, 15--23

\bibitem[{{Maeder}(1992)}]{maeder92}
{Maeder}, A. 1992, in Instabilities in Evolved Super- and Hypergiants, ed.
  C.~{de Jager} \& H.~{Nieuwenhuijzen}, 138

\bibitem[{{Maeder}(2009)}]{maeder_book_2009}
{Maeder}, A. 2009, {Physics, Formation and Evolution of Rotating Stars}

\bibitem[{{Maeder} \& {Conti}(1994)}]{maeder94}
{Maeder}, A. \& {Conti}, P.~S. 1994, \araa, 32, 227

\bibitem[{{Maeder} {et~al.}(2012){Maeder}, {Georgy}, {Meynet}, \&
  {Ekstr{\"o}m}}]{maeder_2012}
{Maeder}, A., {Georgy}, C., {Meynet}, G., \& {Ekstr{\"o}m}, S. 2012, \aap, 539,
  A110

\bibitem[{{Maeder} \& {Meynet}(2000)}]{mm_2000}
{Maeder}, A. \& {Meynet}, G. 2000, \aap, 361, 159

\bibitem[{{Maeder} \& {Meynet}(2010)}]{maeder_2010}
{Maeder}, A. \& {Meynet}, G. 2010, \nar, 54, 32

\bibitem[{{Magic} {et~al.}(2015){Magic}, {Weiss}, \& {Asplund}}]{magic2015}
{Magic}, Z., {Weiss}, A., \& {Asplund}, M. 2015, \aap, 573, A89

\bibitem[{{McEvoy} {et~al.}(2015){McEvoy}, {Dufton}, {Evans}, {Kalari},
  {Markova}, {Sim{\'o}n-D{\'{\i}}az}, {Vink}, {Walborn}, {Crowther}, {de
  Koter}, {de Mink}, {Dunstall}, {H{\'e}nault-Brunet}, {Herrero}, {Langer},
  {Lennon}, {Ma{\'{\i}}z Apell{\'a}niz}, {Najarro}, {Puls}, {Sana},
  {Schneider}, \& {Taylor}}]{mcevoy_2015}
{McEvoy}, C.~M., {Dufton}, P.~L., {Evans}, C.~J., {et~al.} 2015, \aap, 575, A70

\bibitem[{{Meynet}(1992)}]{meynet92}
{Meynet}, G. 1992, in Instabilities in Evolved Super- and Hypergiants, ed.
  C.~{de Jager} \& H.~{Nieuwenhuijzen}, 173

\bibitem[{{Mihalas}(1969)}]{mihalas69}
{Mihalas}, D. 1969, \apjl, 156, L155

\bibitem[{{Moriya} \& {Langer}(2015)}]{moriya_langer_2015}
{Moriya}, T.~J. \& {Langer}, N. 2015, \aap, 573, A18

\bibitem[{{Moriya} {et~al.}(2015){Moriya}, {Sanyal}, \& {Langer}}]{moriya2015}
{Moriya}, T.~J., {Sanyal}, D., \& {Langer}, N. 2015, \aap, 575, L10

\bibitem[{{Muijres} {et~al.}(2011){Muijres}, {de Koter}, {Vink}, {Krti{\v
  c}ka}, {Kub{\'a}t}, \& {Langer}}]{muijres_2011}
{Muijres}, L.~E., {de Koter}, A., {Vink}, J.~S., {et~al.} 2011, \aap, 526, A32

\bibitem[{{Nieuwenhuijzen} \& {de Jager}(1990)}]{nieuwenhuijzen1990}
{Nieuwenhuijzen}, H. \& {de Jager}, C. 1990, \aap, 231, 134

\bibitem[{{Owocki} {et~al.}(2004){Owocki}, {Gayley}, \& {Shaviv}}]{owocki2004}
{Owocki}, S.~P., {Gayley}, K.~G., \& {Shaviv}, N.~J. 2004, \apj, 616, 525

\bibitem[{{Paxton} {et~al.}(2013){Paxton}, {Cantiello}, {Arras}, {Bildsten},
  {Brown}, {Dotter}, {Mankovich}, {Montgomery}, {Stello}, {Timmes}, \&
  {Townsend}}]{mesa2013}
{Paxton}, B., {Cantiello}, M., {Arras}, P., {et~al.} 2013, \apjs, 208, 4

\bibitem[{{Petrovic} {et~al.}(2006){Petrovic}, {Pols}, \&
  {Langer}}]{petrovic_2006}
{Petrovic}, J., {Pols}, O., \& {Langer}, N. 2006, \aap, 450, 219

\bibitem[{{Raskin} {et~al.}(2008){Raskin}, {Scannapieco}, {Rhoads}, \& {Della
  Valle}}]{raskin2008}
{Raskin}, C., {Scannapieco}, E., {Rhoads}, J., \& {Della Valle}, M. 2008, \apj,
  689, 358

\bibitem[{{Ruszkowski} \& {Begelman}(2003)}]{rusz_2003}
{Ruszkowski}, M. \& {Begelman}, M.~C. 2003, \apj, 586, 384

\bibitem[{{Sab{\'{\i}}n-Sanjuli{\'a}n}
  {et~al.}(2014){Sab{\'{\i}}n-Sanjuli{\'a}n}, {Sim{\'o}n-D{\'{\i}}az},
  {Herrero}, {Walborn}, {Puls}, {Ma{\'{\i}}z Apell{\'a}niz}, {Evans}, {Brott},
  {de Koter}, {Garcia}, {Markova}, {Najarro}, {Ram{\'{\i}}rez-Agudelo}, {Sana},
  {Taylor}, \& {Vink}}]{carolina_2014}
{Sab{\'{\i}}n-Sanjuli{\'a}n}, C., {Sim{\'o}n-D{\'{\i}}az}, S., {Herrero}, A.,
  {et~al.} 2014, \aap, 564, A39

\bibitem[{{Saio} {et~al.}(1998){Saio}, {Baker}, \& {Gautschy}}]{saio98}
{Saio}, H., {Baker}, N.~H., \& {Gautschy}, A. 1998, \mnras, 294, 622

\bibitem[{{Sana} {et~al.}(2013){Sana}, {van Boeckel}, {Tramper}, {Ellerbroek},
  {de Koter}, {Kaper}, {Moffat}, {Schnurr}, {Schneider}, \& {Gies}}]{sana_2013}
{Sana}, H., {van Boeckel}, T., {Tramper}, F., {et~al.} 2013, \mnras, 432, L26

\bibitem[{{Sander} {et~al.}(2014){Sander}, {Todt}, {Hainich}, \&
  {Hamann}}]{sander_2014}
{Sander}, A., {Todt}, H., {Hainich}, R., \& {Hamann}, W.-R. 2014, \aap, 563,
  A89

\bibitem[{{Schaerer}(1996)}]{schaerer96}
{Schaerer}, D. 1996, \aap, 309, 129

\bibitem[{{Schneider} {et~al.}(2014){Schneider}, {Izzard}, {de Mink}, {Langer},
  {Stolte}, {de Koter}, {Gvaramadze}, {Hu{\ss}mann}, {Liermann}, \&
  {Sana}}]{schneider_2014}
{Schneider}, F.~R.~N., {Izzard}, R.~G., {de Mink}, S.~E., {et~al.} 2014, \apj,
  780, 117

\bibitem[{{Schnurr} {et~al.}(2008){Schnurr}, {Casoli}, {Chen{\'e}}, {Moffat},
  \& {St-Louis}}]{schnurr_2008}
{Schnurr}, O., {Casoli}, J., {Chen{\'e}}, A.-N., {Moffat}, A.~F.~J., \&
  {St-Louis}, N. 2008, \mnras, 389, L38

\bibitem[{{Schnurr} {et~al.}(2009){Schnurr}, {Moffat}, {Villar-Sbaffi},
  {St-Louis}, \& {Morrell}}]{schnurr_2009}
{Schnurr}, O., {Moffat}, A.~F.~J., {Villar-Sbaffi}, A., {St-Louis}, N., \&
  {Morrell}, N.~I. 2009, \mnras, 395, 823

\bibitem[{{Shaviv}(1998)}]{shaviv1998}
{Shaviv}, N.~J. 1998, \apjl, 494, L193

\bibitem[{{Shaviv}(2001)}]{shaviv2001}
{Shaviv}, N.~J. 2001, \apj, 549, 1093

\bibitem[{{Smith}(2014)}]{smith_2014}
{Smith}, N. 2014, \araa, 52, 487

\bibitem[{{Smith} \& {Arnett}(2014)}]{smith+arnett2014}
{Smith}, N. \& {Arnett}, W.~D. 2014, \apj, 785, 82

\bibitem[{{Smith} {et~al.}(2004){Smith}, {Vink}, \& {de Koter}}]{smith2004}
{Smith}, N., {Vink}, J.~S., \& {de Koter}, A. 2004, \apj, 615, 475

\bibitem[{{Soderberg} {et~al.}(2008){Soderberg}, {Berger}, {Page}, {Schady},
  {Parrent}, {Pooley}, {Wang}, {Ofek}, {Cucchiara}, {Rau}, {Waxman}, {Simon},
  {Bock}, {Milne}, {Page}, {Barentine}, {Barthelmy}, {Beardmore}, {Bietenholz},
  {Brown}, {Burrows}, {Burrows}, {Byrngelson}, {Cenko}, {Chandra}, {Cummings},
  {Fox}, {Gal-Yam}, {Gehrels}, {Immler}, {Kasliwal}, {Kong}, {Krimm},
  {Kulkarni}, {Maccarone}, {M{\'e}sz{\'a}ros}, {Nakar}, {O'Brien}, {Overzier},
  {de Pasquale}, {Racusin}, {Rea}, \& {York}}]{soderberg2008}
{Soderberg}, A.~M., {Berger}, E., {Page}, K.~L., {et~al.} 2008, \nat, 453, 469

\bibitem[{{Spruit}(2002)}]{spruit02}
{Spruit}, H.~C. 2002, \aap, 381, 923

\bibitem[{{Stothers} \& {Chin}(1973)}]{stot_chin1973}
{Stothers}, R. \& {Chin}, C.-W. 1973, \apj, 179, 555

\bibitem[{{Stothers}(2003)}]{stothers_2003}
{Stothers}, R.~B. 2003, \apj, 589, 960

\bibitem[{{Stothers} \& {Chin}(1993)}]{stot_chin1993}
{Stothers}, R.~B. \& {Chin}, C.-W. 1993, \apjl, 408, L85

\bibitem[{{Suijs} {et~al.}(2008){Suijs}, {Langer}, {Poelarends}, {Yoon},
  {Heger}, \& {Herwig}}]{suijs2008}
{Suijs}, M.~P.~L., {Langer}, N., {Poelarends}, A.-J., {et~al.} 2008, \aap, 481,
  L87

\bibitem[{{Taylor} {et~al.}(2011){Taylor}, {Evans}, {Sana}, {Walborn}, {de
  Mink}, {Stroud}, {Alvarez-Candal}, {Barb{\'a}}, {Bestenlehner}, {Bonanos},
  {Brott}, {Crowther}, {de Koter}, {Friedrich}, {Gr{\"a}fener},
  {H{\'e}nault-Brunet}, {Herrero}, {Kaper}, {Langer}, {Lennon}, {Ma{\'{\i}}z
  Apell{\'a}niz}, {Markova}, {Morrell}, {Monaco}, \& {Vink}}]{taylor_2011}
{Taylor}, W.~D., {Evans}, C.~J., {Sana}, H., {et~al.} 2011, \aap, 530, L10

\bibitem[{{Trampedach} {et~al.}(2014){Trampedach}, {Stein},
  {Christensen-Dalsgaard}, {Nordlund}, \& {Asplund}}]{trampedach2014}
{Trampedach}, R., {Stein}, R.~F., {Christensen-Dalsgaard}, J., {Nordlund},
  {\AA}., \& {Asplund}, M. 2014, \mnras, 445, 4366

\bibitem[{{Ulmer} \& {Fitzpatrick}(1998)}]{ulmer_fitz98}
{Ulmer}, A. \& {Fitzpatrick}, E.~L. 1998, \apj, 504, 200

\bibitem[{{van Marle} {et~al.}(2008){van Marle}, {Owocki}, \&
  {Shaviv}}]{van_marle_2008}
{van Marle}, A.~J., {Owocki}, S.~P., \& {Shaviv}, N.~J. 2008, \mnras, 389, 1353

\bibitem[{{Vink} {et~al.}(2000){Vink}, {de Koter}, \& {Lamers}}]{vink2000}
{Vink}, J.~S., {de Koter}, A., \& {Lamers}, H.~J.~G.~L.~M. 2000, \aap, 362, 295

\bibitem[{{Vink} {et~al.}(2001){Vink}, {de Koter}, \& {Lamers}}]{vink2001}
{Vink}, J.~S., {de Koter}, A., \& {Lamers}, H.~J.~G.~L.~M. 2001, \aap, 369, 574

\bibitem[{{Wentzel}(1970)}]{wentzel1970}
{Wentzel}, D.~G. 1970, \apj, 160, 373

\bibitem[{{Yoon} {et~al.}(2006){Yoon}, {Langer}, \& {Norman}}]{yln06}
{Yoon}, S., {Langer}, N., \& {Norman}, C. 2006, \aap, 460, 199

\bibitem[{{Yorke}(1986)}]{yorke_1986}
{Yorke}, H.~W. 1986, \araa, 24, 49

\bibitem[{{Yusof} {et~al.}(2013){Yusof}, {Hirschi}, {Meynet}, {Crowther},
  {Ekstr{\"o}m}, {Frischknecht}, {Georgy}, {Abu Kassim}, \&
  {Schnurr}}]{yusof2013}
{Yusof}, N., {Hirschi}, R., {Meynet}, G., {et~al.} 2013, \mnras, 433, 1114

\bibitem[{{Zickgraf} {et~al.}(1985){Zickgraf}, {Wolf}, {Stahl}, {Leitherer}, \&
  {Klare}}]{zickgraf_1985}
{Zickgraf}, F.-J., {Wolf}, B., {Stahl}, O., {Leitherer}, C., \& {Klare}, G.
  1985, \aap, 143, 421

\end{thebibliography}


\clearpage
\begin{appendix}
\section{Interior structure of a $85\mathrm{M_{\sun}}$ stellar model}
\label{app:inflations_eg}
   \begin{figure}
\centering
\resizebox{\hsize}{!}{\includegraphics[angle=-90]{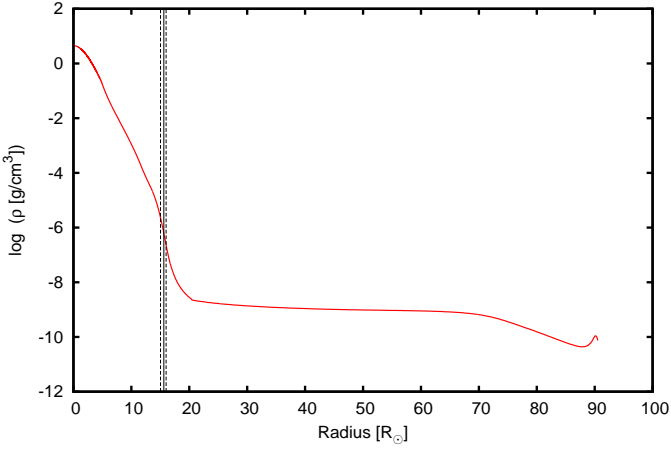}}
   \caption[]{Density structure of the stellar model. The black solid line marks the 
   base of the inflated envelope, i.e. where $\beta=0.15$. The intersection of the dotted lines with 
   the red line on either side mark the points where $\beta=0.15\pm 0.045$.} 
\label{fig:app:dens}
\end{figure}

   \begin{figure}
\centering	
\resizebox{\hsize}{!}{\includegraphics[angle=-90]{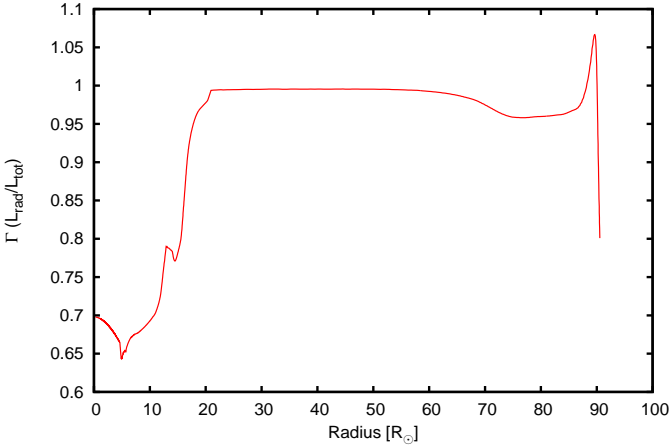}}
   \caption[]{Run of $\Gamma$ in the interior of the stellar model.}
\label{fig:app:gamma}
\end{figure}

   \begin{figure}
\centering	
\resizebox{\hsize}{!}{\includegraphics[angle=-90]{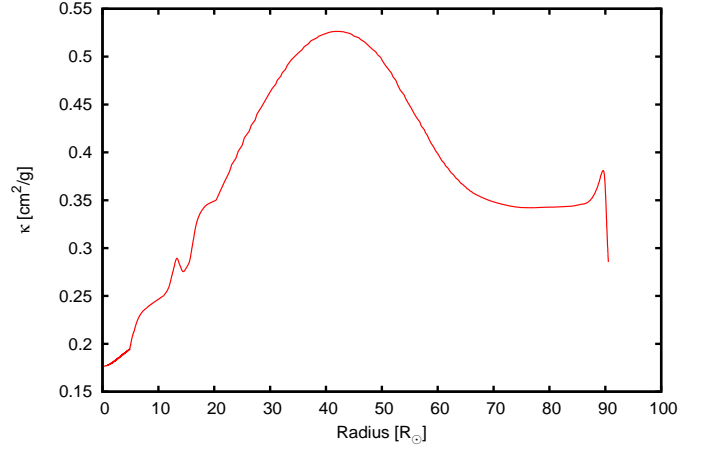}}
   \caption[]{Rosseland mean opacity $\kappa$ in the interior of the stellar model.}
\end{figure}

   \begin{figure}
\centering
\resizebox{\hsize}{!}{\includegraphics[angle=-90]{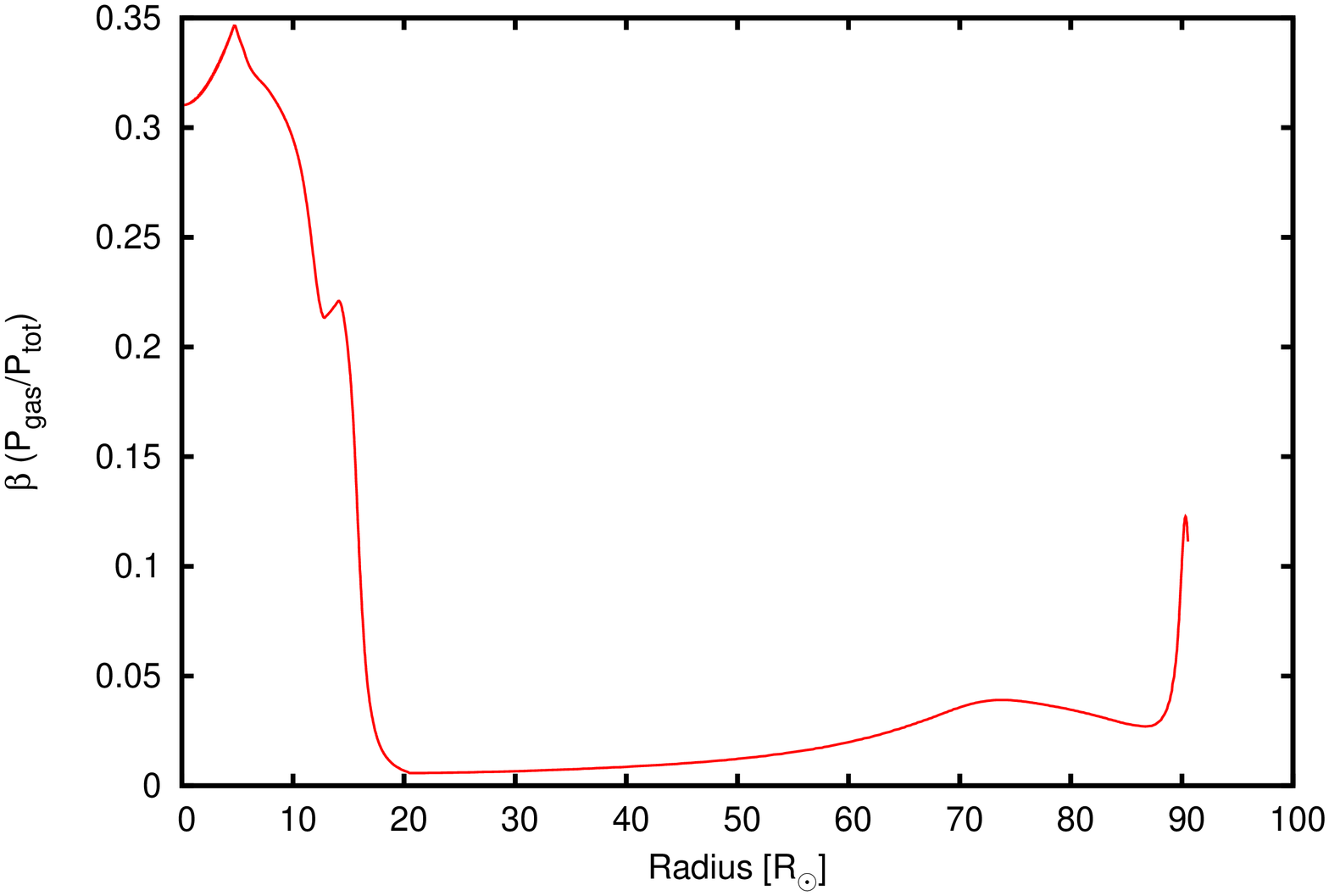}}
   \caption[]{Run of $\beta(=P_\mathrm{{gas}}/P_{\mathrm{tot}})$ in the interior of the stellar model.}
   \label{fig:app:beta}
\end{figure}

   \begin{figure}
\centering	
\resizebox{\hsize}{!}{\includegraphics[angle=-90]{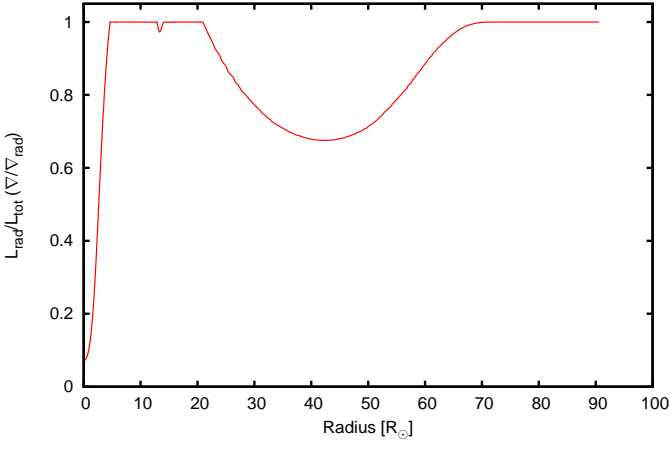}}
   \caption[]{Fraction of flux carried by radiation ($L_{\mathrm{rad}}/L_{\mathrm{tot}}$) in the interior of the stellar model.}
\end{figure}

\section{Effect of efficient convection on inflation}
\label{app:mix_length_comparison}
\begin{figure}
\centering
\resizebox{\hsize}{!}{\includegraphics[angle=-90]{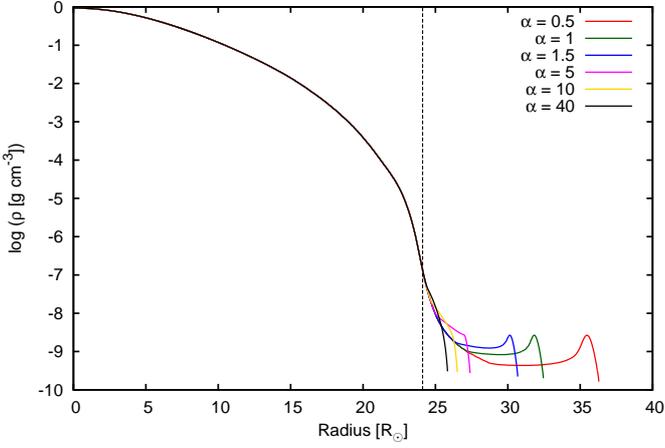}}
   \caption[]{Density profile of a $300\,\mso$ model with different values of the mixing 
   length parameter $\alpha$ (see Sec.\,\ref{sec:modelling}). The black dotted line marks the 
   location of $r_{\rm core}$, i.e. the base of the inflated envelope where $\beta = 0.15$. }
\label{fig:app:mix_length_comparison}
\end{figure}

Knowing that convective flux is proportional to the mixing length, we 
show here (Fig.\,\ref{fig:app:mix_length_comparison}) that by increasing the mixing length parameter $\alpha$ in an inflated $300\mso$ model near the ZAMS, 
the inflation gradually goes away and what we are left with is an almost non-inflated star, whose radius is
well-approximated by core radius $r_{\rm core}$ of the inflated model. 

\section{Convective velocity profile in a WR model}
\label{app:supersonic}
\begin{figure}
\centering
\resizebox{\hsize}{!}{\includegraphics[angle=-90]{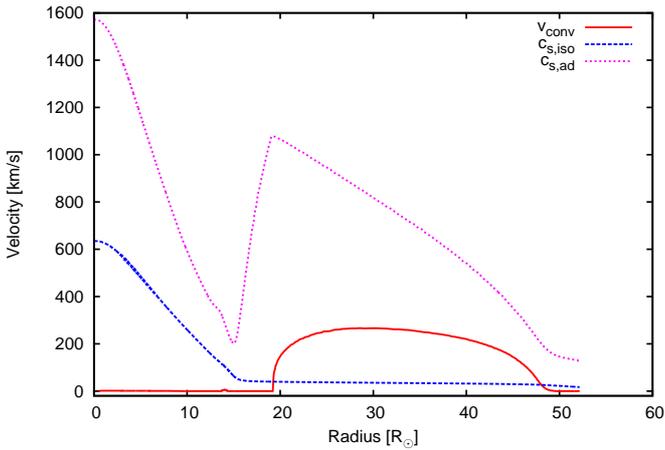}}
   \caption[]{Convective velocity, isothermal sound speed and adiabatic sound speed profiles 
   in a  $147\,\mathrm{M_{\sun}}$ WNL type star with $Y_{s}=89\%$. }
\label{fig:app:conv_vel_eg}

\end{figure}

The convective velocity is shown as a function of radius in a massive 
($147\,\mathrm{M_{\sun}}$) WR-type ($Y_S=0.89$) stellar 
model, in Fig.\,\ref{fig:app:conv_vel_eg}. The variation of isothermal and adiabatic sound speeds are also plotted for comparison. 
The convective velocities exceed the local isothermal sound speed in the envelope where 
conditions are non-adiabatic, i.e. thermal adjustment time is short. In such models, turbulent pressure 
becomes important (which is not taken into account in our models) as well as standard MLT fails 
to be a good approximation for modelling convection.

\section{Representative models}
\label{app:example_models}
The profiles of different relevant physical quantities are shown for a 
few selected stellar models at five distinct effective temperatures corresponding to 
the three peaks in $\Gamma_{\deb{max}}$ and the two troughs in between the peaks (cf., Fig.\,\ref{fig:gamma_HR}).

\begin{figure*}
\centering
\resizebox{\hsize}{!}{\includegraphics[angle=-90]{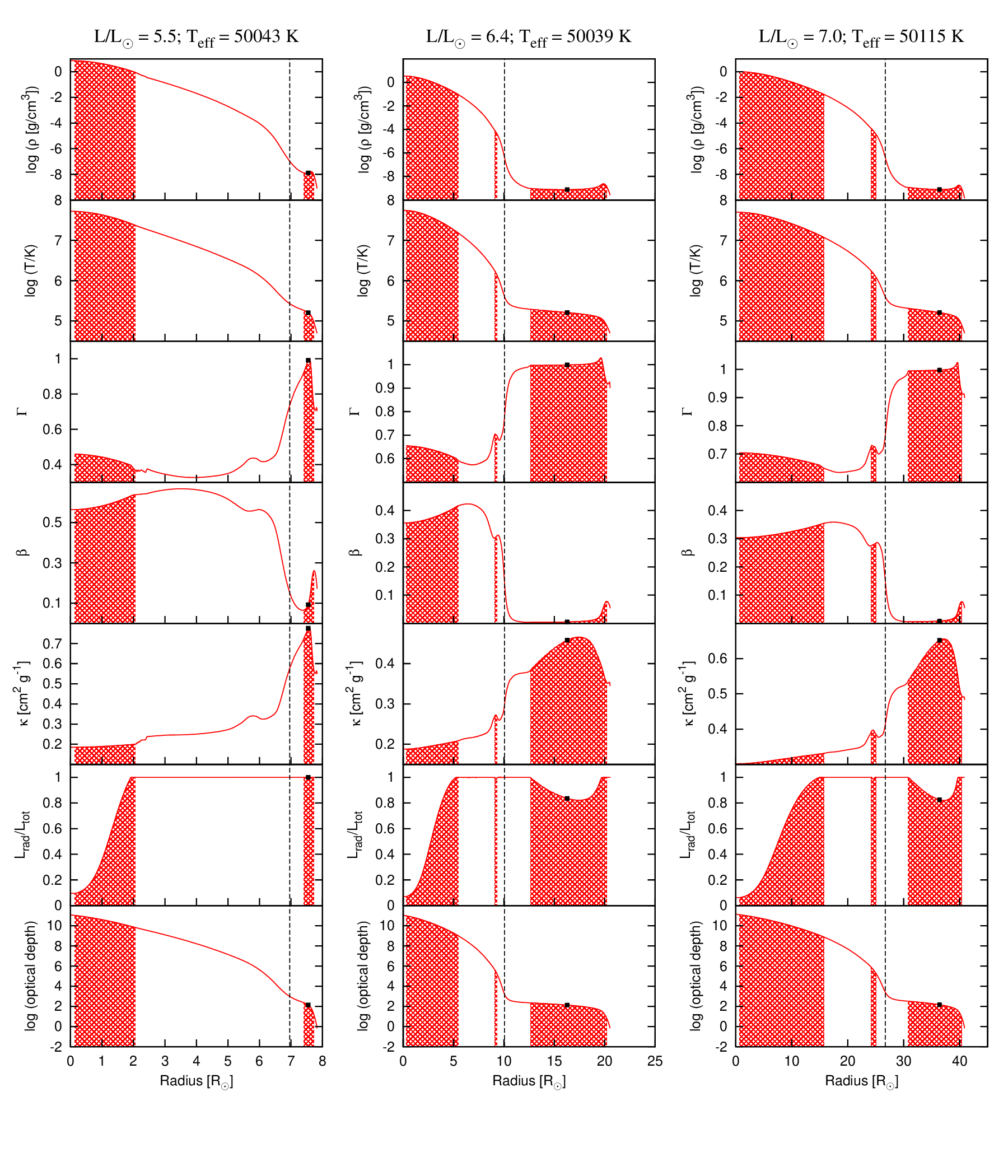}}
   \caption[]{Detailed structure examples for stellar models with an effective temperature near 50\,000\,K,
   for three different luminosities (cf., Fig.\,2). 
   The dashed line marks the point at which 
   $\beta$ falls below $0.15$, i.e. the beginning of the inflated envelope. The square symbol 
   marks the temperature $T_{\deb{Fe}}$ at which $\kappa$ is maximum due to the iron opacity bump. The hatched regions show the convective zones.}
\label{fig:app:Febump}
\end{figure*}

\begin{figure*}
\centering
\resizebox{\hsize}{!}{\includegraphics[angle=-90]{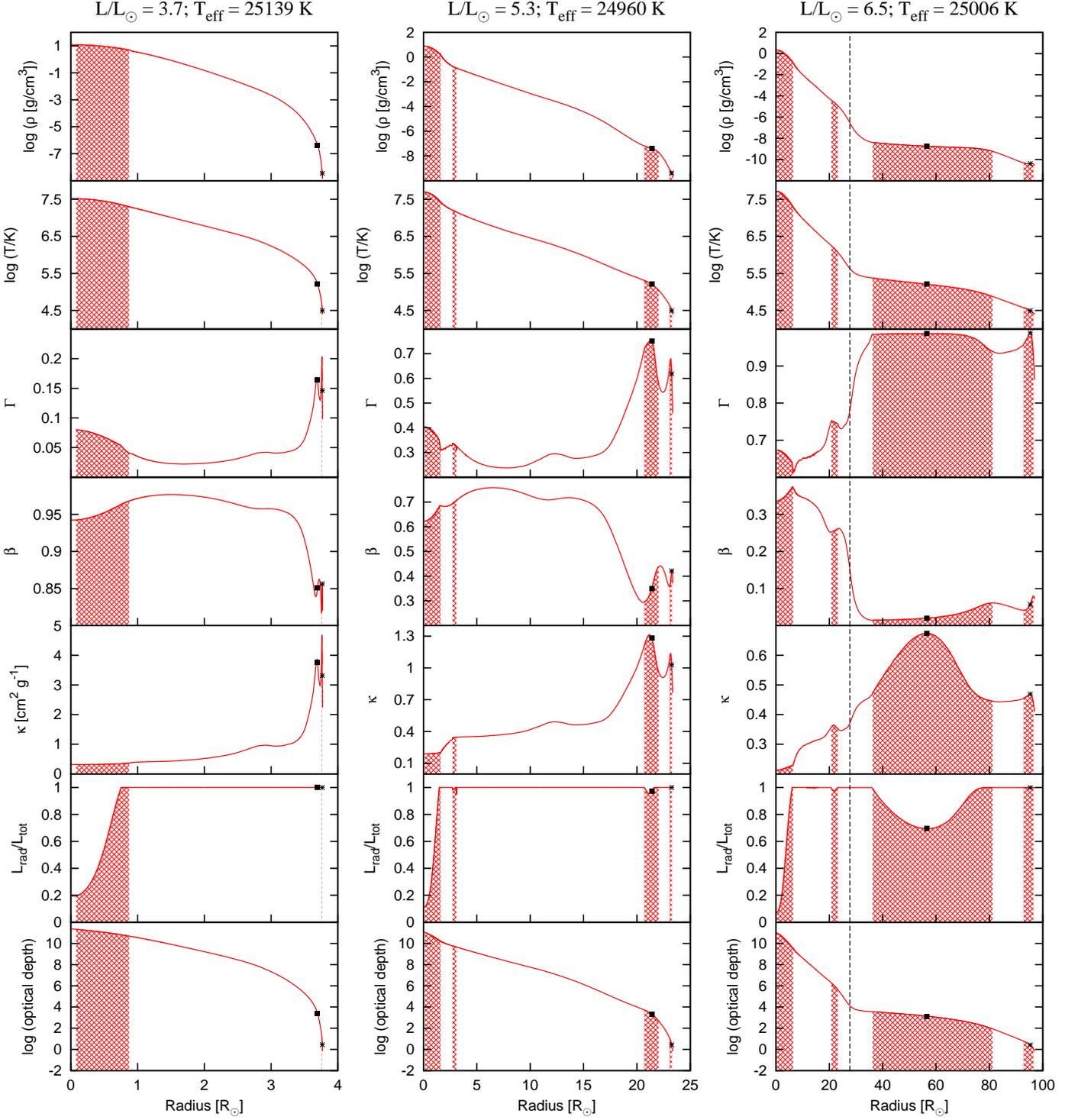}}
   \caption[]{Detailed structure examples for stellar models with an effective temperature near 25\,000\,K, 
   for three different luminosities (cf., Fig.\,2). 
   The dashed line marks the point at which 
   $\beta$ falls below $0.15$, i.e. the beginning of the inflated envelope. The square and the cross 
   mark the temperatures $T_{\deb{Fe}}$ and $T_{\deb{Fe}}$ at which $\kappa$ 
   is maximum due to the iron and the helium opacity bumps respectively. The hatched regions show the convective zones.}
\label{fig:app:Hebump}
\end{figure*}

\begin{figure*}
\centering
\resizebox{\hsize}{!}{\includegraphics[angle=-90]{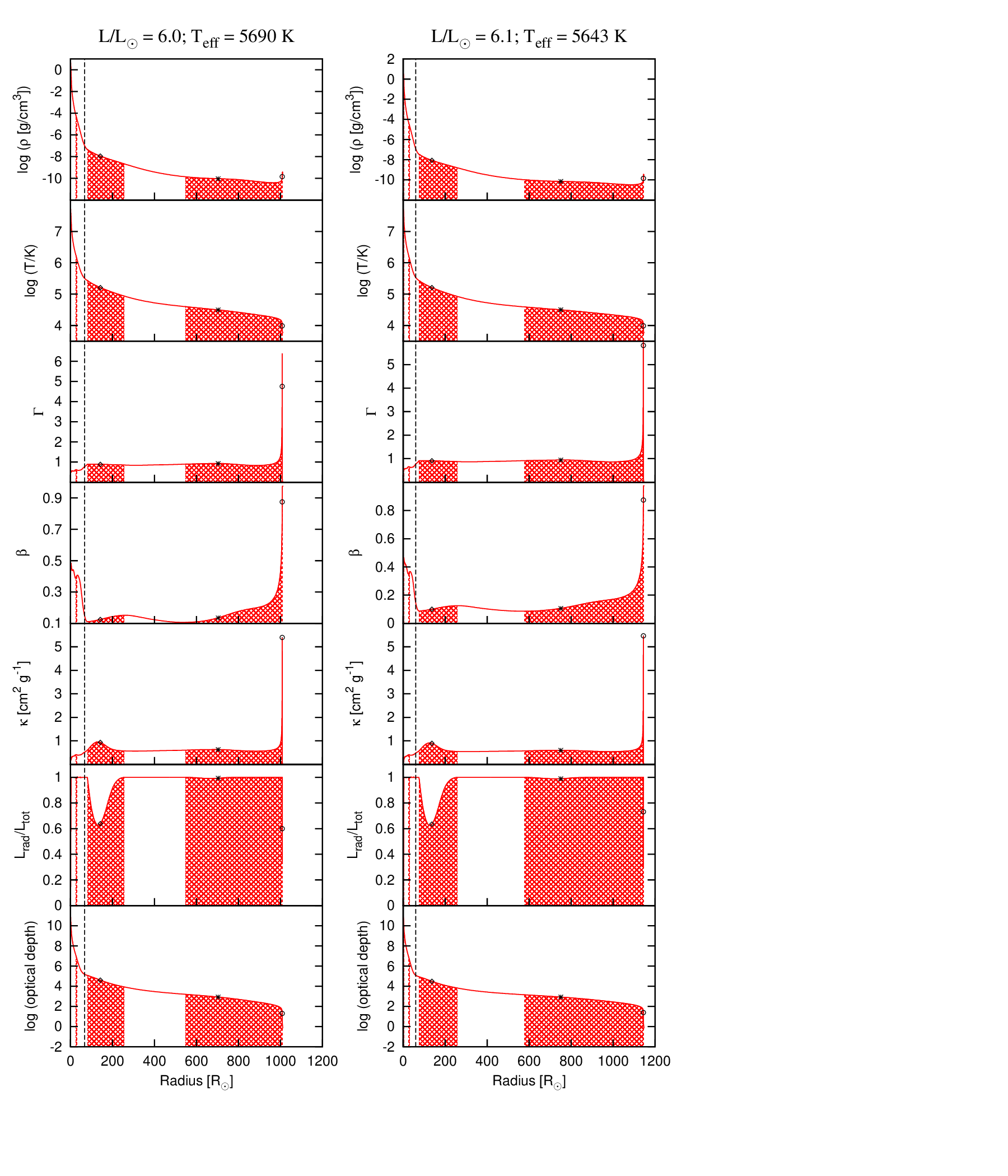}}
\label{fig:app:H-bump}
   \caption[]{Detailed structure examples for stellar models with an effective temperature near 5\,000\,K, 
   for two different luminosities (cf., Fig.\,2).
   The dashed line marks the point at which 
   $\beta$ falls below $0.15$, i.e. the beginning of the inflated envelope. The square, cross and the circle   
   mark the temperatures $T_{\deb{Fe}}$, $T_{\deb{Fe}}$ and $T_{\deb{H}}$ at which $\kappa$ 
   is maximum due to the iron, helium and hydrogen opacity bumps respectively. The hatched regions show the convective zones.}
\label{fig:app:Hbump}
\end{figure*}

\begin{figure*}
\centering
\resizebox{\hsize}{!}{\includegraphics[angle=-90]{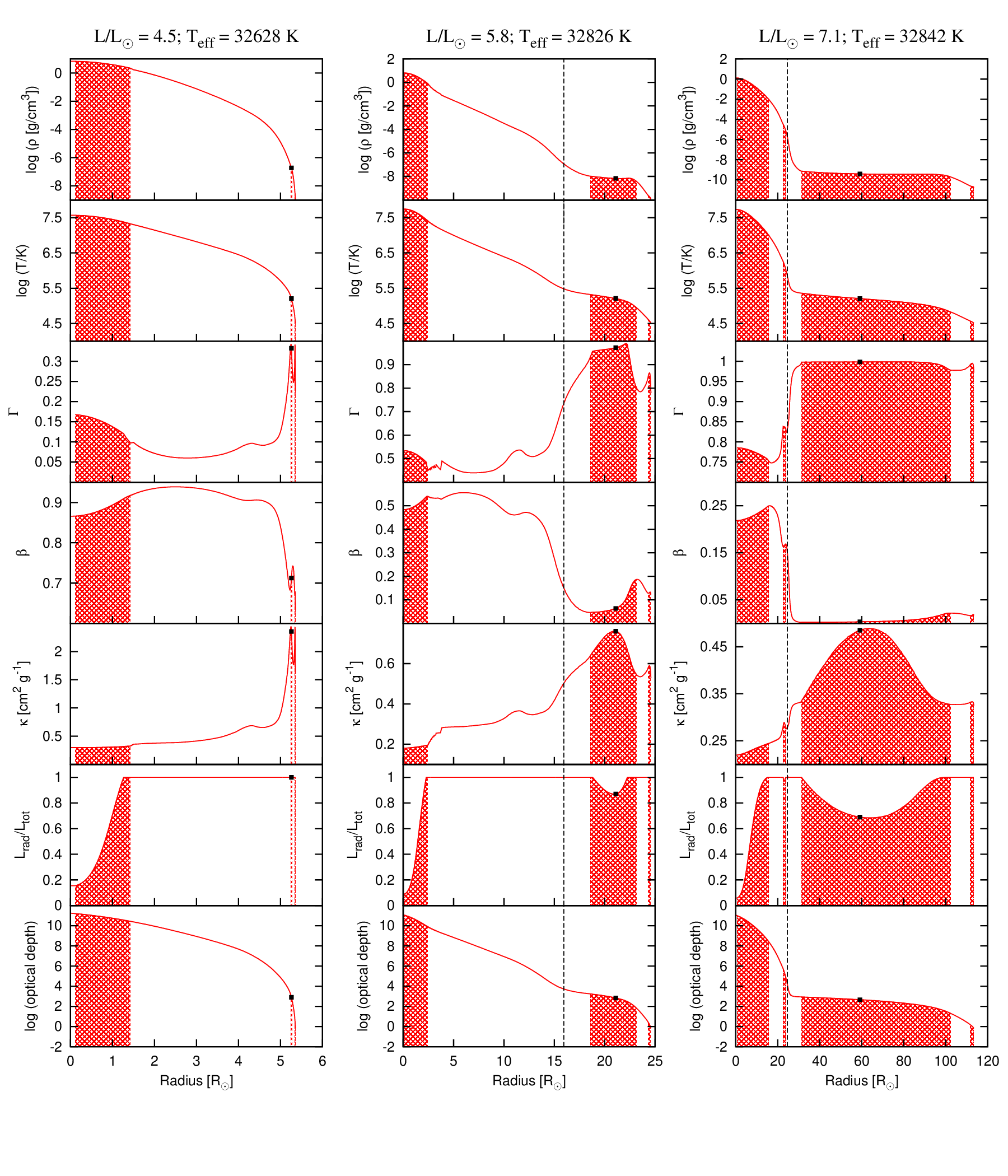}}
   \caption[]{Detailed structure examples for stellar models with an effective temperature near 32\,000\,K, 
   for three different luminosities (cf., Fig.\,2). 
   The dashed line marks the point at which 
   $\beta$ falls below $0.15$, i.e. the beginning of the inflated envelope. The square symbol 
   marks the temperature $T_{\deb{Fe}}$ at which $\kappa$ is maximum due to the iron opacity bump. The hatched regions show the convective zones.}
\label{fig:app:Fe-Hebump}
\end{figure*}

\begin{figure*}
\centering
\resizebox{\hsize}{!}{\includegraphics[angle=-90]{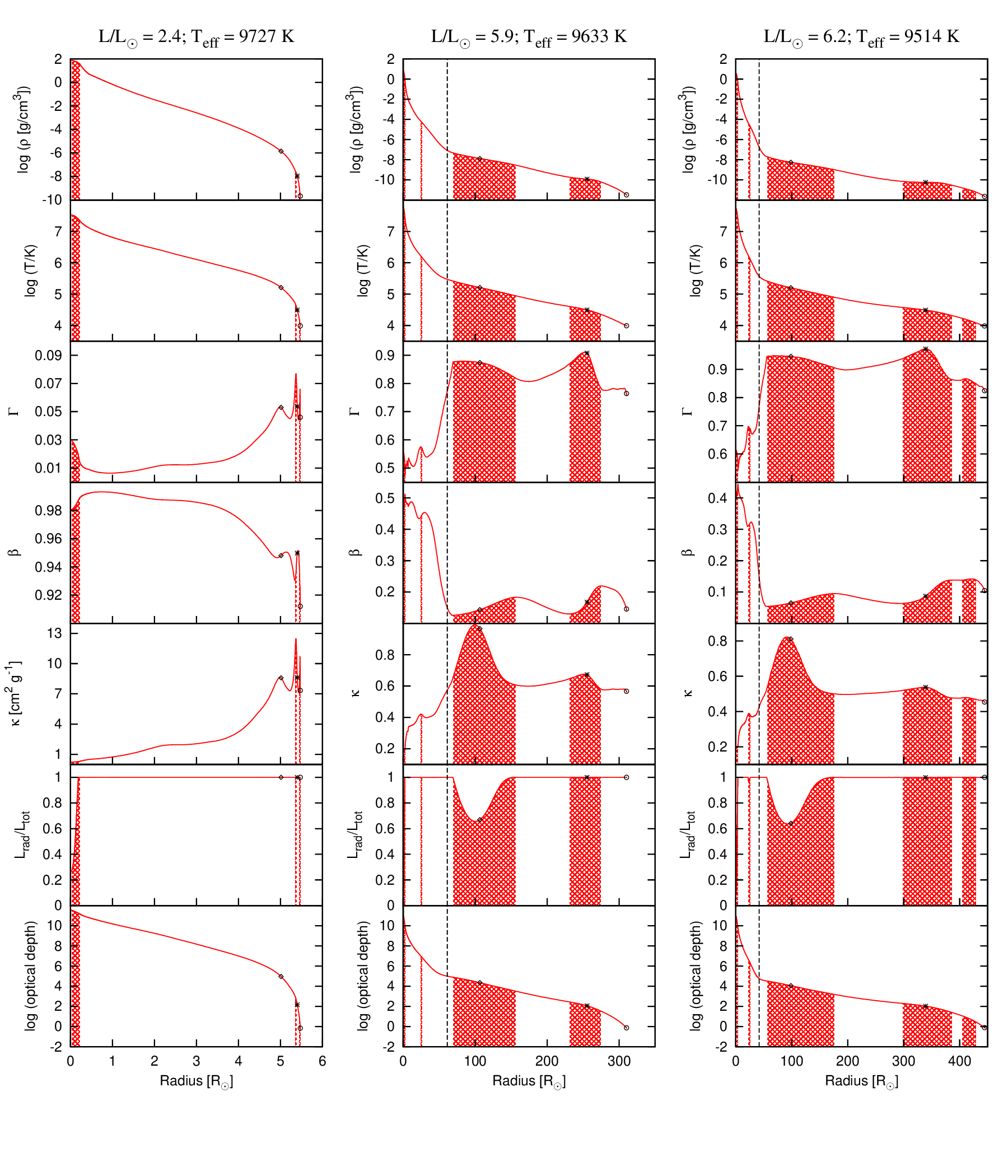}}
   \caption[]{Detailed structure examples for stellar models with an effective temperature near 10\,000\,K, 
   for three different luminosities (cf., Fig.\,2). 
   The dashed line marks the point at which 
   $\beta$ falls below $0.15$, i.e. the beginning of the inflated envelope. The square, cross and the circle   
   mark the temperatures $T_{\deb{Fe}}$, $T_{\deb{Fe}}$ and $T_{\deb{H}}$ at which $\kappa$ 
   is maximum due to the iron, helium and hydrogen opacity bumps respectively. The hatched regions show the convective zones.}
\label{fig:app:He-Hbump}
\end{figure*}

\end{appendix}

\end{document}